\documentclass[12pt]{article}
\usepackage{amsmath,amsfonts}
\usepackage{amscd,amsthm}
\usepackage{geometry}
\usepackage{graphicx}
\usepackage{color}
\usepackage[applemac]{inputenc}
\usepackage{tikz}
\usepackage{graphicx}
\usepackage{color}
\usepackage[applemac]{inputenc}
\usepackage{cases}
\usepackage{tikz}
\usepackage{algorithm2e}
\usepackage{setspace} 
\usepackage{hyperref}

\begin{document}

\title{The Miracle of Peer Review and Development in Science: An Agent-Based Model\thanks{The authors gratefully acknowledge support from the "Lend\"{u}let" Program of the Hungarian Academy of Sciences and from the Hungarian Scientific Research Fund (OTKA K 112929). K\'{a}roly Tak\'{a}cs acknowledges funding from the European Research Council (ERC) under the European Union's Horizon 2020 research and innovation programme (grant agreement No 648693).}}
\author{Simone Righi \thanks{Department of Agricultural Sciences and Technologies, Alma Mater Studiorum - University of Bologna, Mailing Address: Viale Fanin 50, 40127, Bologna, Italy
Email: s.righi@unibo.it} \thanks{MTA TK "Lend\"{u}let" Research Center for Educational and Network Studies (RECENS),
Hungarian Academy of Sciences. Mailing address:
Orsz\'{a}gh\'{a}z utca 30, 
1014 Budapest, Hungary.
Email: simone.righi@tk.mta.hu .} \and K\'{a}roly Tak\'{a}cs \thanks{MTA TK "Lend\"{u}let" Research Center for Educational and Network Studies (RECENS),
Hungarian Academy of Sciences. Mailing address:
Orsz\'{a}gh\'{a}z utca 30, 
1014 Budapest, Hungary.
Email: takacs.karoly@tk.mta.hu}}

\date{\today}  
\maketitle

\begin{abstract}
It is not easy to rationalize how peer review, as the current grassroots of science, can work based on voluntary contributions of reviewers. There is no rationale to write impartial and thorough evaluations. Consequently, there is no risk in submitting low-quality work by authors. As a result, scientists face a social dilemma: if everyone acts according to his or her own self-interest, low scientific quality is produced. Still, in practice, reviewers as well as authors invest high effort in reviews and submissions. 
 We examine how the increased relevance of public good benefits (journal impact factor), the editorial policy of handling incoming reviews, and the acceptance decisions 
that take into account reputational information can help the evolution of high-quality contributions from authors. High effort from the side of reviewers is problematic even if authors cooperate: reviewers are still best off by producing low-quality reviews, which does not hinder scientific development, just adds random noise and unnecessary costs to it. We show with agent-based simulations that tacit agreements between authors that are based on reciprocity might decrease these costs, but does not result in superior scientific quality. Our study underlines why certain self-emerged current practices, such as the increased importance of journal metrics, the reputation-based selection of reviewers, and the reputation bias in acceptance
work efficiently for scientific development. Our results find no answers, however, how the system of peer review with impartial and thorough evaluations could be sustainable jointly with rapid scientific development.

\vspace{0.4cm}

\end{abstract}

\section{Introduction}

Peer review is the fundamental process used by the scientific community to ensure the quality of academic publications (cf. e.g., \cite{Alberts2008,Bornmann2013,SquazzoniTakacs2011}). Several generations of scientists have contributed high-quality reviews, while only authorship has been credited for academic career. It is not easy to rationalize why researchers provide impartial reviews and constructive advice voluntarily, as they need to sacrifice time that could be used for their own research activities \cite{Bernstein2013}. In consequence, it is a puzzle how the system of peer review can be sustainable at all.

This puzzle can be described as a double social dilemma game \cite{BianchiEtAlWp,SquazzoniBravoTakacs2013} where scientists can choose levels of efforts for both manuscripts and reviews. Given the presence of costs in terms of time and effort, no contribution (sloppy review) is the best reply strategy for reviews. When all scientists play according to their best reply strategy, the resulting outcome is no scientific control on the quality of submitted papers. The dominant strategy equilibrium of low quality reviews mean that submissions need not be of high quality. Due to the immense costs of producing high quality work, authors are best off by submitting poor manuscripts. In short, in the lack of explicit sanctions and incentives, low-quality submissions and low-quality reviews from the dominant strategy equilibrium in the social dilemma of scientific production. 

Although social dilemmas of this kind are difficult to resolve in general, certain theoretical solutions have already been proposed that can be applied to the context of peer review. The most evident improvement might come from shifting the payoffs in favor of cooperation. For instance, the reward for overall cooperation can be introduced by attaching higher importance on scientific quality through the introduction or  increased emphasis on journal metrics, such as impact factor. Another straightforward solution is the application of selective incentives \cite{Olson1965}: allocating additional benefits for authors of high-quality papers (e.g., promotion and grants, which is current practice) and for reviewers writing high-quality reports (which is rare in current practice).

Direct \cite{Axelrod1984,Axelrod1981} and indirect reciprocity \cite{Boyd1989,Milinski2002_1,Milinski_2002a,Milinski_2001,Nowak2006,Nowak2005,Semmann2005} are potential solutions to social dilemmas, if the chance of repetition is high and the opportunities for retaliation are in foresight. In the context of peer review, indirect reciprocity can be facilitated, for instance, by rotating the roles of authors, reviewers, and editors \cite{BravoSquazzoniTakacs} . The social embeddedness of scientific production further enhances the chance of solving the social dilemma efficiently \cite{Barrera2008,Coleman1986}. The small world aspect of working in specific fields, the intertwined network of co-authorship, participation in international project consortia, and hangouts at conferences all reduce the competitiveness and improve the pro-social character of reviewing. 

Furthermore, once informal communication, gossip, reputation, image scoring, and stratification enters into the structure of the social dilemma, cooperation might emerge and be maintained \cite{Sommerfeld2007,Sommerfeld}. Recording, keeping, and relying on reputational scores has become an efficient guideline for cooperation in many areas of life \cite{Dellarocas2003}.
Earlier achievement and reputational information certainly plays an important role in current practice both for editorial and reviewer decisions \cite{Paolucci2014}.
 
Social incentives that are associated with direct and indirect reciprocity and the structural embeddedness of peer review might be key mechanisms that rationalize cooperative behavior of reviewers. The importance of social incentives is highlighted also by surveys asking for the motivations of reviewers \cite{Malicki,Warne2016}. In fact, monetary incentives might be in conflict with or drive out social incentives when applied to reviewers \cite{Chetty2014,SquazzoniBravoTakacs2013}. 
These solutions for social dilemmas might help us to understand how peer review can work and be sustained. Once the fundamental mechanisms are studied rigorously, they can also lead to policy recommendations on improvements of the current system and the design of new solutions (cf. \cite{PaolucciGrimaldo2014}).

The rest of this paper is divided in three sections. In Section \ref{themodel}, we  introduce our agent-based model of scientific work and peer review. We report simulations results in Section \ref{results} and we conclude in Section \ref{conclusion}.

\section{The model}
\label{themodel}

\subsection{The baseline model of peer review}

We model the production of scientific work and its peer review by a simple agent-based model. In this paper, we build on the view that the aim of peer review is to ensure scientific production and evaluation \cite{Gilbert1997}. The model emphasizes the costly character of high-quality manuscript submissions and reviewer contributions. We take it as granted that high-quality submissions lead to better science, but reviews impact scientific quality only indirectly. Our model makes radical simplifications on the practical aspects of the peer review system intentionally. This way, we aim at providing a straightforward assessment of the institutional conditions and editorial policies under which authors are motivated to produce high-quality submissions and reviewers are motivated to provide high quality reviews. Hence, we are primarily interested in the emergence of cooperation that results in scientific quality.

The model contains $N$ scientists as agents who write single-author papers. At each discrete period $t$, each scientist performs the task of an author - by producing one article -  and the task of a reviewer. Authors and reviewers comprise of an identical set of agents. For the sake of simplicity, we consider a single journal with a single editor, who is not an author or a reviewer in the journal.
Authors can produce low or high quality contributions that they submit to the editor. The editor selects $\mu$ (set to two in our simulations) reviewers for each paper choosen uniformly at random\footnote{An alternative specification, where editor decision are proportional to the reputation of scientists has been studied with results qualitatively similar to those proposed in the next session.}  with an upper limit $k$ of reviews for each reviewer (set to four for the sake of our simulations). We assume that reviewers always accept requests, but they do not necessarily invest high effort in performing reviews. It is of their strategic decision to produce a low quality review at low cost (normalized to zero) or a high quality review at high cost. The former is the best reply strategy of the reviewer if no further incentives are provided. The review is translated into a binary recommendation (accept or reject) which is passed on to the editor. We assume that the recommendation is random with fifty-fifty percent chance of accept or reject in case of low reviewer effort and reflects the quality of the submission perfectly in case of high reviewer effort.
The editor's decision is based on incoming recommendations and it has a binary outcome: accept or reject. We consider a single round of review. Acceptance benefits the author, but benefits the editor (the journal and scientific development in general) only if the submission was of high quality.

The incentive structure and the strategy space of the game are defined as follows. The editor wants to maximize high-quality scientific output in the journal and would like to minimize the number of low-quality articles appearing in the journal. This means that similarly to the model of \cite{BianchiEtAlWp}, papers that are accepted and of bad quality are the most harmful for the journal. These values are used as output measures to evaluate the performance of the entire system in our simulations. 

The situation is of asymmetric information \cite{SquazzoniBravoTakacs2013}, in which the editor is unable to assess the true quality of submissions or the true effort of reviewers. The true quality of accepted papers is revealed probabilistically after publication (that we will fix to $P_{reveal}=1$ in the simulations reported) and there is no way for the editor to assess the true quality of rejected submissions. The crucial parts of editorial policy are therefore the selection of reviewers who can be trusted for their recommendations and the extent of reliance on reviewer recommendations. Several strategies could possibly be used in order to arrive at proper conclusion. We vary these editorial strategies in between simulations, because these
could be easily translated into policy recommendations.

Keeping a reputational account of scientists is part of all such strategies. Editorial reputations of scientists are improved largely by high quality publications (named $GP$), degraded even more by low quality publications ($BP$), worsened by rejected submissions ($Rej$), improved by good reviews ($GR$) and worsened by bad reviews ($BR$). Good reviews are those where the paper is revealed of the same quality of what the reviewer said. Bad reviews are those where there is a difference, i.e. where the paper is revealed as bad while the reviewer said that it was good or where the paper was revealed as good, but the reviewer recommended rejection (in case of conflicting reviews). In summary, the editorial reputation of scientist $i$ is given as:

\begin{equation}
REP_i^E = \#GP-\#Rej- \alpha*\#BP+ \gamma*(\#GR-\#BR)
\end{equation}

where $\#Rej=\#BN+\#GN$ (the sum of bad and good rejected papers) and $0<\gamma<1$. Please note the asymmetric character of reviewer bias: as the high quality of rejected papers is never revealed, rejection is a safer strategy for reviewers (a rejection recommendation can only turn out to be a bad review in case other reviewers recommended publication, the paper has been published, and its true high quality has been revealed). Parameter $\alpha>1$ represents the relative detrimental effect for the journal reputation of accepting a low quality paper. 

Incoming reviews are assumed to lead to editorial conclusions according to an editorial policy which is assumed to be fixed and not updated within a simulation. We manipulate editorial strategies, between the simulations in order to compare the effectiveness of these policies. Editorial strategies differ with regard to reviewer selection, the handling of conflicting reviews, and relying on author reputations in desk rejection and acceptance. The last element will only be added to the extended model. We consider four editorial strategies with regard to reviewer selection and handling of conflicting reviews. For all of them, if all referees agree, then the editor follows the unanimous advice. In case of \textit{disagreement}, the editor can use one of the following strategies:
\begin{itemize}
\item \textbf{AP}: Reject;
\item \textbf{1P}: Accept;
\item \textbf{ER}: Follow the advice of one of the referees chosen at random  probability proportional to the relative editorial reputations of the referees;
\item \textbf{MR}: Follow the advice of the most reputed referee.
\end{itemize}


Authors have perfect information about the quality of their own submissions. Similarly to the model of \cite{BianchiEtAlWp}, authors decide to submit a paper of low quality (at low cost) or at high quality (at high cost). They are best off with the publication of their low-quality papers. Thus, in terms of obtained payoffs: $BP > GP > BN > GN$.
For the simulations we assume that all agents receive a unit of endowment $e+E$ (research time) in each period, which they lose by investing high effort in writing a high-quality paper ($e$) and high-quality reviews ($E$, where $0<E<e$). We assume no gain from not being published ($V_{N}=0$). Being published yields a positive return $V_{P}>e$. The numerical payoffs in our simulations are:

\begin{eqnarray}
\pi_{BP} = V_{P} + e = 2 + 1 = 3 \\
\pi_{GP} = V_{P} = 2 \\
\pi_{BN} = V_{N} + e = 0 + 1 = 1 \\
\pi_{GN} = V_{N} = 0 \\
\end{eqnarray}
 
When in the role of reviewers, scientists first accept editorial requests to review submissions. We assume that this does not entail any significant costs: the time spent on pushing the "accept" button in an editorial managerial system is negligible and the social costs of being committed to reviewing are counterbalanced by gaining access to papers before their publication. 
Once papers are assigned to reviewers, they decide to invest low effort (no loss from the endowment of valuable time) or high effort in performing the task. 
A high reviewing effort implies a deduction from the endowment $E$ that is proportional to the ratio between number of good reviews and the number of assigned reviews. 
Low effort is the best reply strategy of reviewers. 

Altogether, writing high quality papers as well as writing high quality reviews entails sacrificing valuable research time for scientists. For the sake of simplicity, we assume that endowments are not transferable between the two kind of activities. In this way, unlike \cite{BianchiEtAlWp}, we disregard the potential time conflict between writing high-quality papers and reviews. In their model, scientists need to allocate time between submissions and reviewing. In our model, both activities are costly if they are performed well.

In line with the current duality of practices, two cases are compared: single blind and double blind reviews \cite{Seeber2016}. In case of single blind reviews, reviewers can condition their recommendations on the reputation of the author. In case of double blind reviews, only the editor is able to make decisions based on the reputation of authors. In case of a single blind review system, the reviewer strategy can be conditional on the public reputation of the author. Unlike the editorial reputation of authors, the public reputation of an author $i$ depends only on \textit{published} papers and does not severely punishes bad publications:

\begin{equation}
REP_i^P = \#GP-\#BP
\end{equation}


 As reviewers are also authors, author and reviewer strategies are bundled and are characterized by the following elements:
\begin{itemize}
\item A decision for the production of manuscripts can be either: 
\begin{itemize}
\item $c$: produce manuscripts of high quality (cooperation) 
\item $d$: produce manuscripts of low quality (defection)
\end{itemize}
\item A decision for the production of reviews that can be either: 
\begin{itemize}
\item $C$: produce reviews at high effort (cooperation)
\item $D$: produce reviews at low effort (defection)
\item $Rep$: exercise high effort with a likelihood that is proportional to the public reputation of the author of the paper (cooperation conditional on reputation)\footnote{This option is only available for the single blind case.}.
\end{itemize}
\end{itemize}

Individual strategies are assigned to scientists at the start randomly and in equal numbers for each combination. This results in four bundled strategies for the double blind case and in six bundled strategies for the single blind case. Authors decide according to their strategy types to submit a paper of low quality or at high quality (at high cost), which is sent out for review according to the rules determined by the editorial policy. Reviewers act according to their strategies and provide recommendations of accept or reject to the editor. Papers are selected for publication as a result of the recommendations of the reviewers and the editorial policy. 

The evolution of author/reviewer strategies is modeled with a replicator dynamics rule adjusted to a finite population. Scientists adopt strategies that ensured higher average payoffs in the previous time period in the population, while disregard strategies that resulted in lower payoffs. Specifically, at a given time $t$ each individual has a probability $P_{evo}$ of being selected for updating his strategy. If this happens, then his new strategy is selected randomly with a probability proportional to the difference of the given strategy to the average expected payoff, weighted for the current strategy frequencies. Formally, the expected new population size for strategy $j$ is given by: 

\begin{equation}
N^j_{t+1}= N^j_t + \left(\overline{P_t^j} \cdot N^j_t - (N_t^j-\overline{P_t})\right)\delta
\end{equation}
where $\overline{P_t^j}$ is the average payoff of strategy $j$, while $\overline{P_t}$ is the average payoff in the whole population and $\delta$ is the speed of evolution. We constrain the finite replicator dynamics process such that all strategies have an integer number of representations in the population and no $N^j_{t+1}<0$. 
Note that the strategies of agents evolve and not the agents are replaced. This means that agents accumulate editorial and public reputation throughout the simulation.

\subsection{Reputation-based selection and Journal Impact Factor}
 
In order to keep the model simple and to concentrate on some key mechanisms, other details are fixed to some natural values and some possible elements can be activated upon choice, which we list below.

\paragraph{Limit to the number of publications.} 
The number of publications is limited to a fixed proportion $\epsilon \ll 1$ of submissions. If the referee process produces too many accepted papers given the current editorial policy of the editor, then all accepted contributions are ranked according to editorial reputation $REP_i^E$ of the authors and only the first $\epsilon N$ papers are published. Throughout the paper, for simulations reported, we consider a medium level of competition with $\epsilon=0.3$.

\paragraph{Journal Impact factor.}
In recognition of the fact that publishing in a reputed journal produces a payoff that depends also on the quality of past published papers, agents who publish a paper may receive an increase to their payoff equal to:
\begin{equation}
JIF_t=\kappa*\frac{\sum GP_t}{\sum GP_t + \sum BP_t}
\end{equation}

At each given time step $t$, the public good benefit of Journal Impact Factor (JIF) is given to all authors who get their papers published, irrespective of quality. The higher the proportion of high-quality papers, the higher the JIF is. Note that the introduction of JIF increases payoffs for publications produced at high effort, but increases free rider rewards for those who are able to publish low-quality work to the same extent. We assume that the Journal Impact Factor is a public good for those who published with a linear production function, where the increment $\kappa \geq 1$ describes the public value of a single scientific contribution. The lack of a Journal Impact Factor and a linear public good with $\kappa \leq 1$ describes a situation that is worse than the Prisoner's Dilemma: even full cooperation does not compensate for the entailed costs ($R \leq P$). Considering the Journal Impact Factor and $\kappa>1$, translates the game into a true linear Public Good Game, which is still extremely difficult to solve. For the sake of the simulations of this paper we assume $\kappa=2$. 

\paragraph{Desk-rejections and speeding up publication.}
With increased time-pressure and burden of reviewers, it is common practice that not all submissions are sent out for review. Some submissions are desk-rejected and others have an easy, speedy route for publication. Editorial decisions behind are very much based on the reputation of authors. We implement desk-rejection and desk-acceptance as an additive feature compared to the baseline model.

Desk-rejection and acceptance introduces a bias in favour of individuals with higher reputation and in damage of individuals with lower editorial reputation compared to the baseline model. Essentially, we introduce the possibility for the editor to desk-reject or accept submissions proportional to the relative editorial reputation of the author, regardless of recommendations by referees. Please note that as producing low or high-quality reviews is also part of editorial reputations, in this way, reviewers receive some compensation for their low or high reviewing efforts.
The rule added to editorial decisions is as follows:
\begin{itemize}
\item Compute the minimal editorial reputation $min(REP_i^E)$, the maximal editorial reputation $max(REP_i^E)$ and the median editorial reputation $median(REP_i^E)$ of agents.
\item If $REP_i^E<median(REP_i^E)$, then with probability
\begin{equation}
P_{da}=\frac{REP_i^E - min(REP_i^E)}{median(REP_i^E)-min(REP_i^E)}
\label{deskacceptance} 
\end{equation}
the paper is desk rejected, and with probability $1- P_{da}$, it is sent to the referees.
\item If $REP_i^E>median(REP_i^E)$, then with probability:
\begin{equation}
P_{dr}=\frac{REP_i^E - median(REP_i^E)}{max(REP_i^E)-median(REP_i^E)}
\label{deskrejection}
\end{equation}
 the paper is desk accepted, and with probability $1- P_{dr}$, it is sent out for review.
\end{itemize}

\section{Results}
\label{results}

\subsection{Baseline results}

\begin{figure}
\centering
\includegraphics[width=0.24\textwidth]{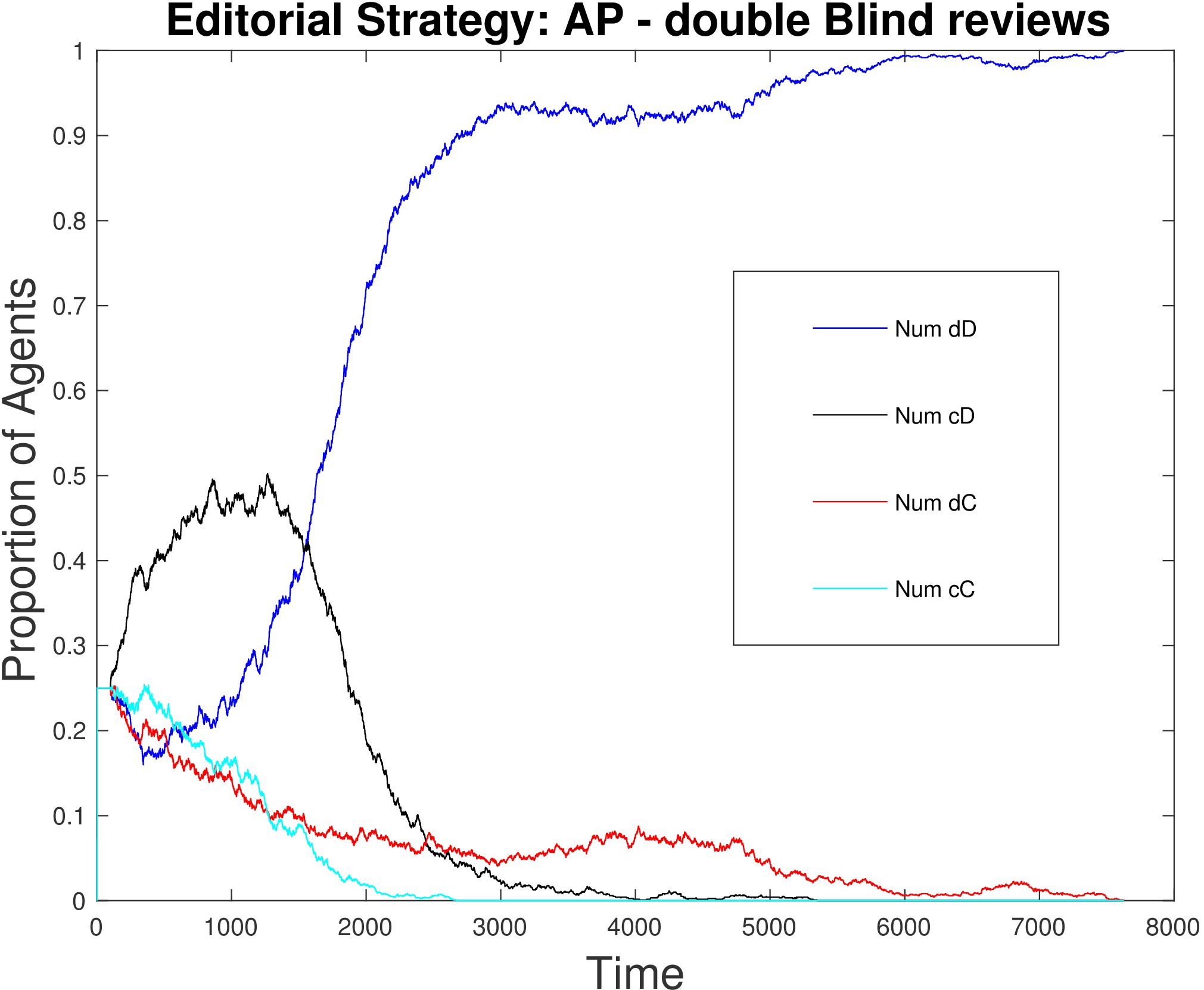}
\includegraphics[width=0.24\textwidth]{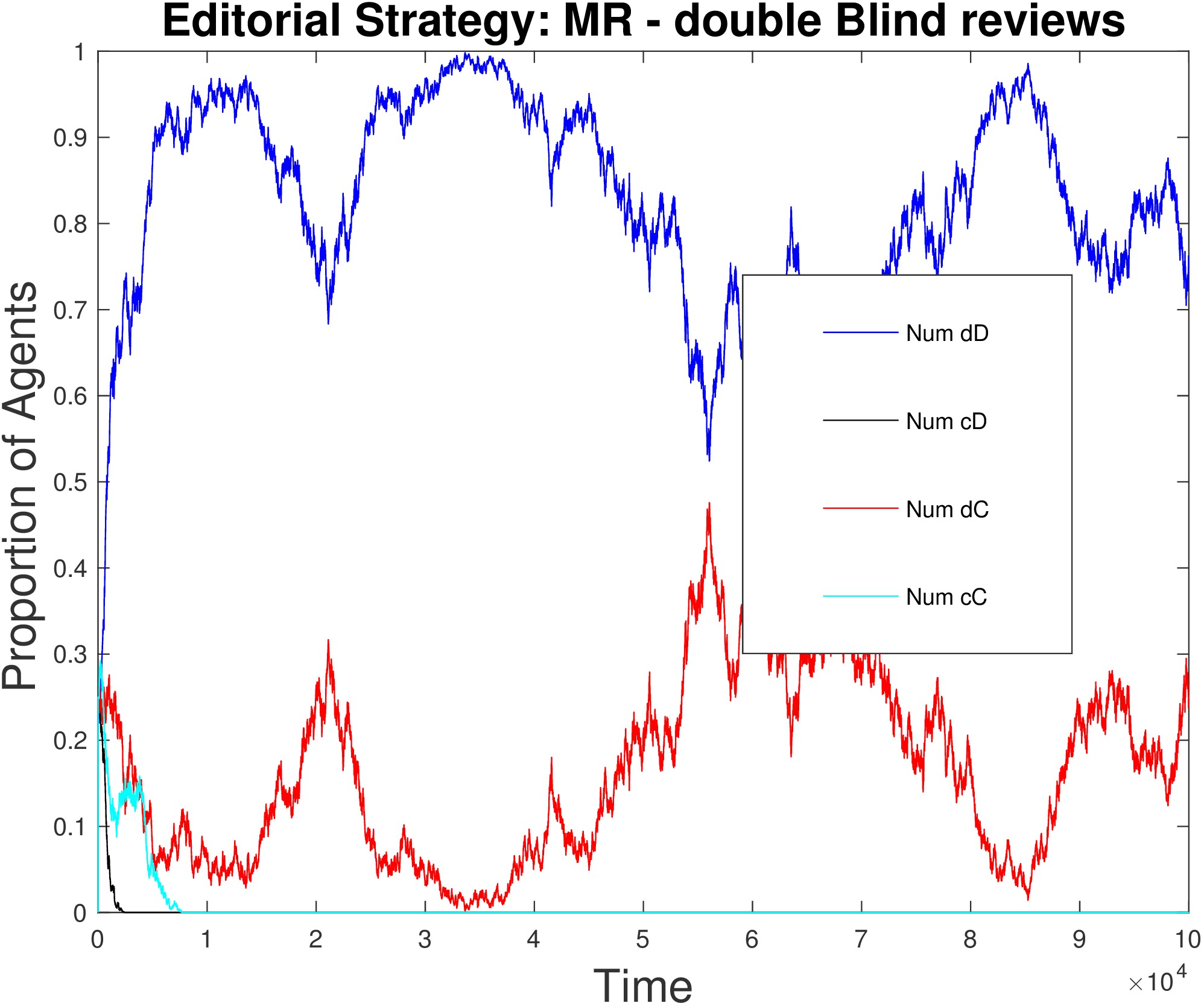}
\includegraphics[width=0.24\textwidth]{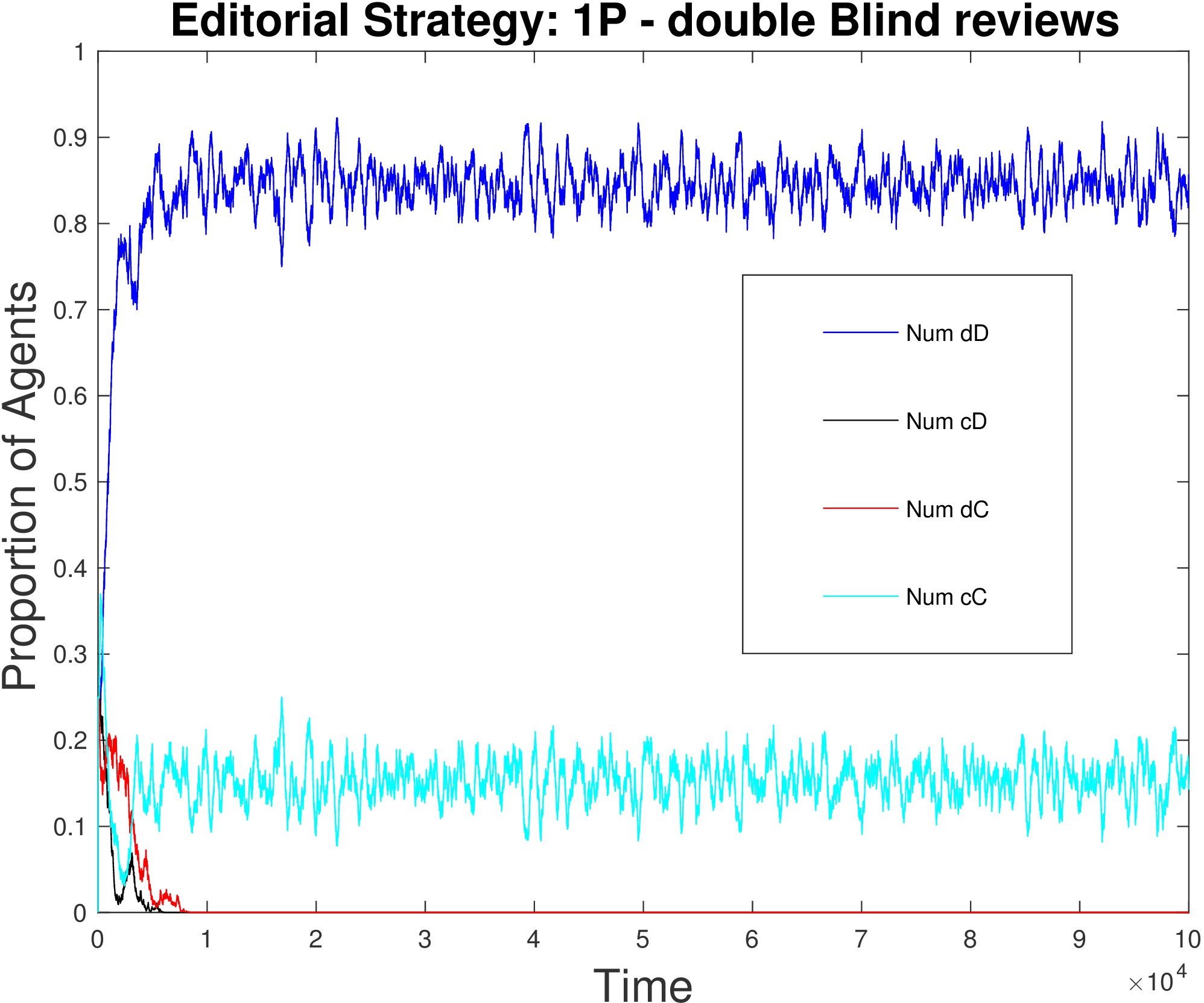}
\includegraphics[width=0.24\textwidth]{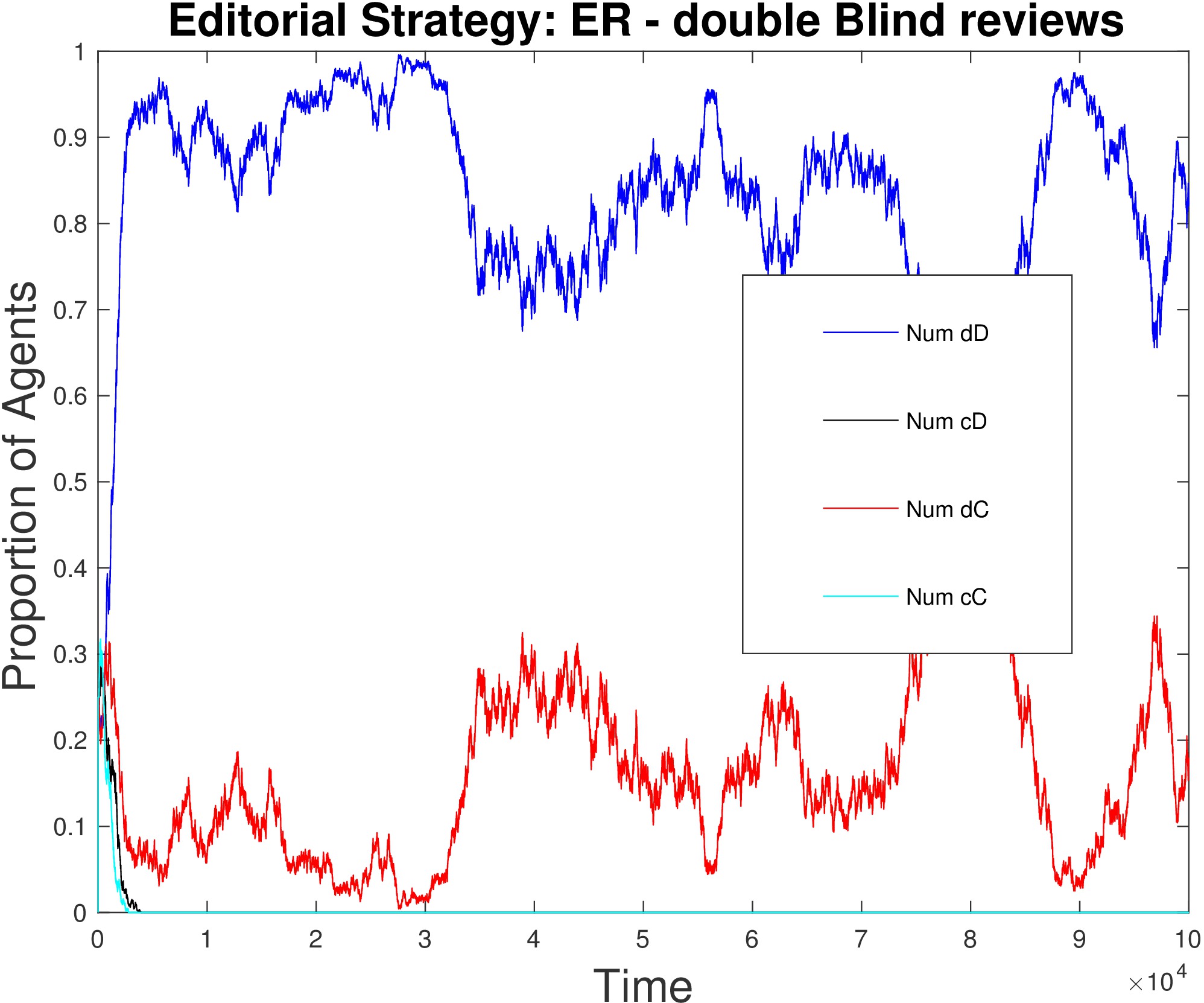}
\caption{The failure of scientific cooperation in the baseline model with double blind reviews (2 random referees) under different editorial policies. Respectively: AP (First Panel), MR (Second Panel), 1P (Third Panel), ER (Fourth Panel). All illustrative simulations concern the baseline model: the population is composed of 1200 individuals; initially divided among the four possible bundled strategies; simulations run till convergence or until $t_{max}=100000$; papers are assigned in random order to randomly chosen referees; each referee can review up to 4 papers and a maximum of  30\% of the papers are publishable in each period.}
\label{Baseline_doubleblind}
\end{figure}

\begin{figure}
\centering
\includegraphics[width=0.24\textwidth]{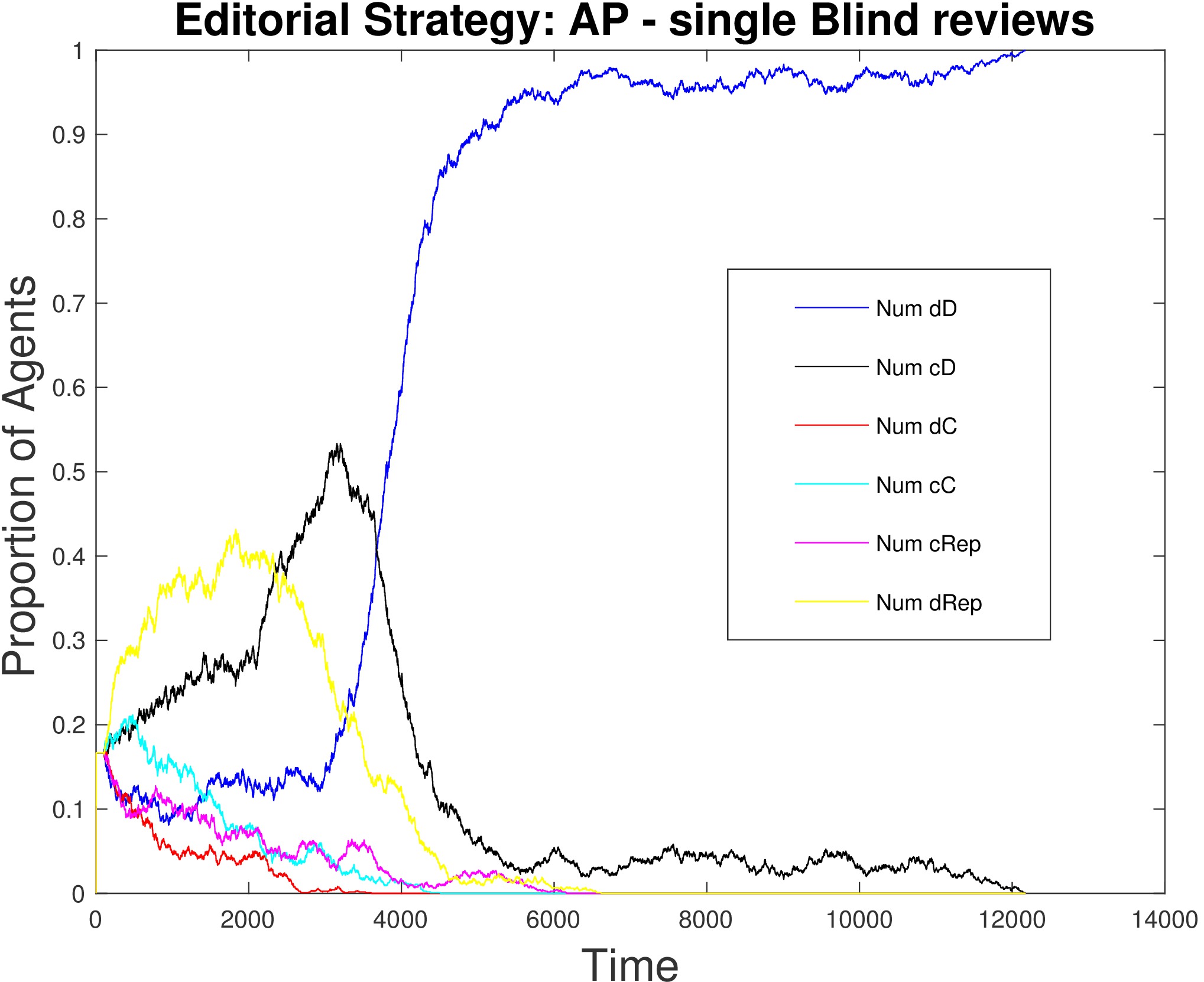}
\includegraphics[width=0.24\textwidth]{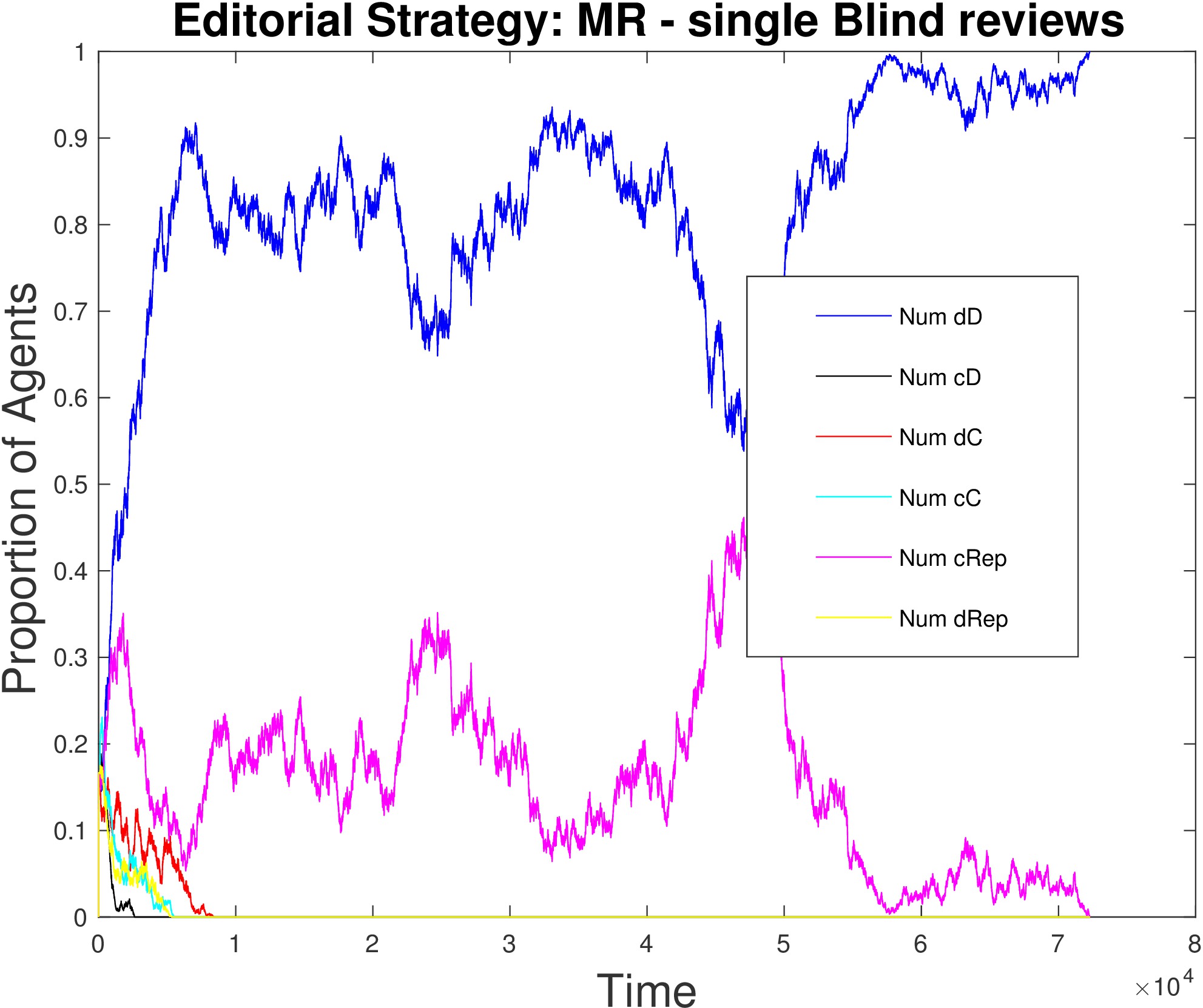}
\includegraphics[width=0.24\textwidth]{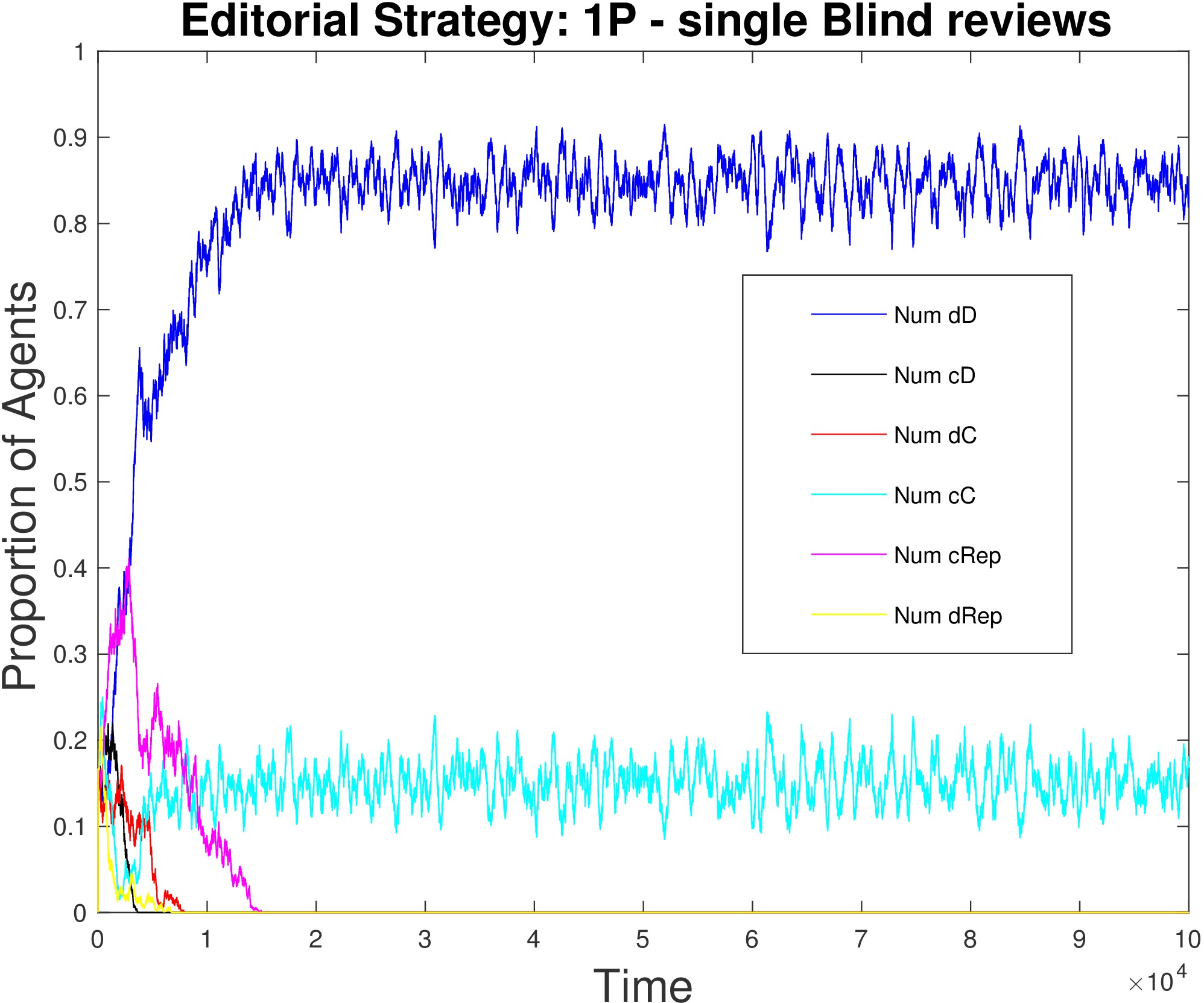}
\includegraphics[width=0.25\textwidth]{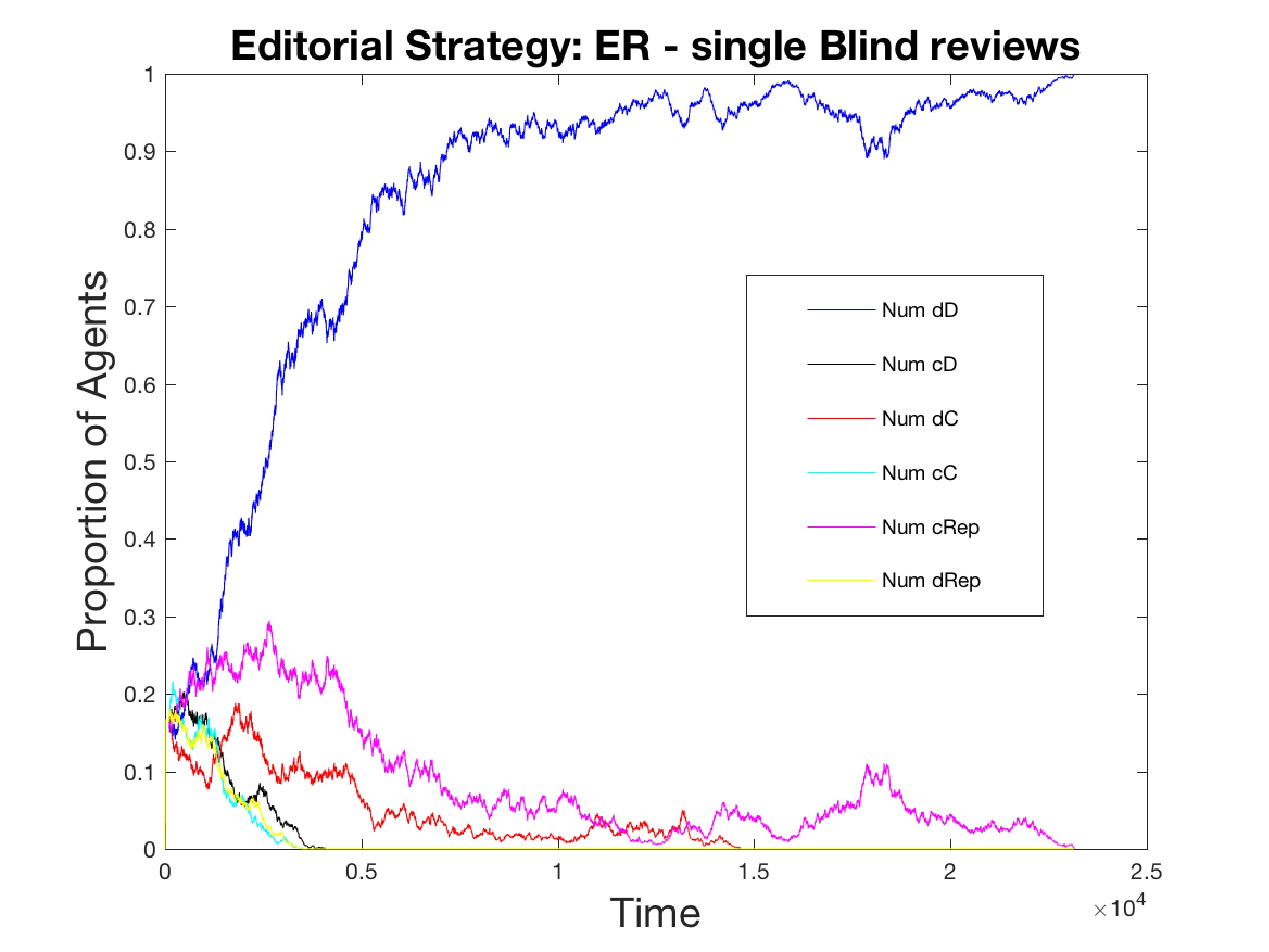}
\caption{The failure of scientific cooperation in the baseline model with single blind reviews (2 random referees) under different editorial policies. Respectively: AP (First Panel), MR (Second Panel), 1P (Third Panel), ER (Fourth Panel). All illustrative simulations concern the baseline model: the population is composed of 1200 individuals initially divided among the 6 possible bundled strategies; simulations run till convergence or until $t_{max}=100000$; papers are assigned in random order to randomly chosen referees; each referee can review up to 4 papers and a maximum of  30\% of the papers are publishable in each period.}
\label{Baseline_singleblind}
\end{figure}

\begin{figure}
\centering
\includegraphics[width=5cm]{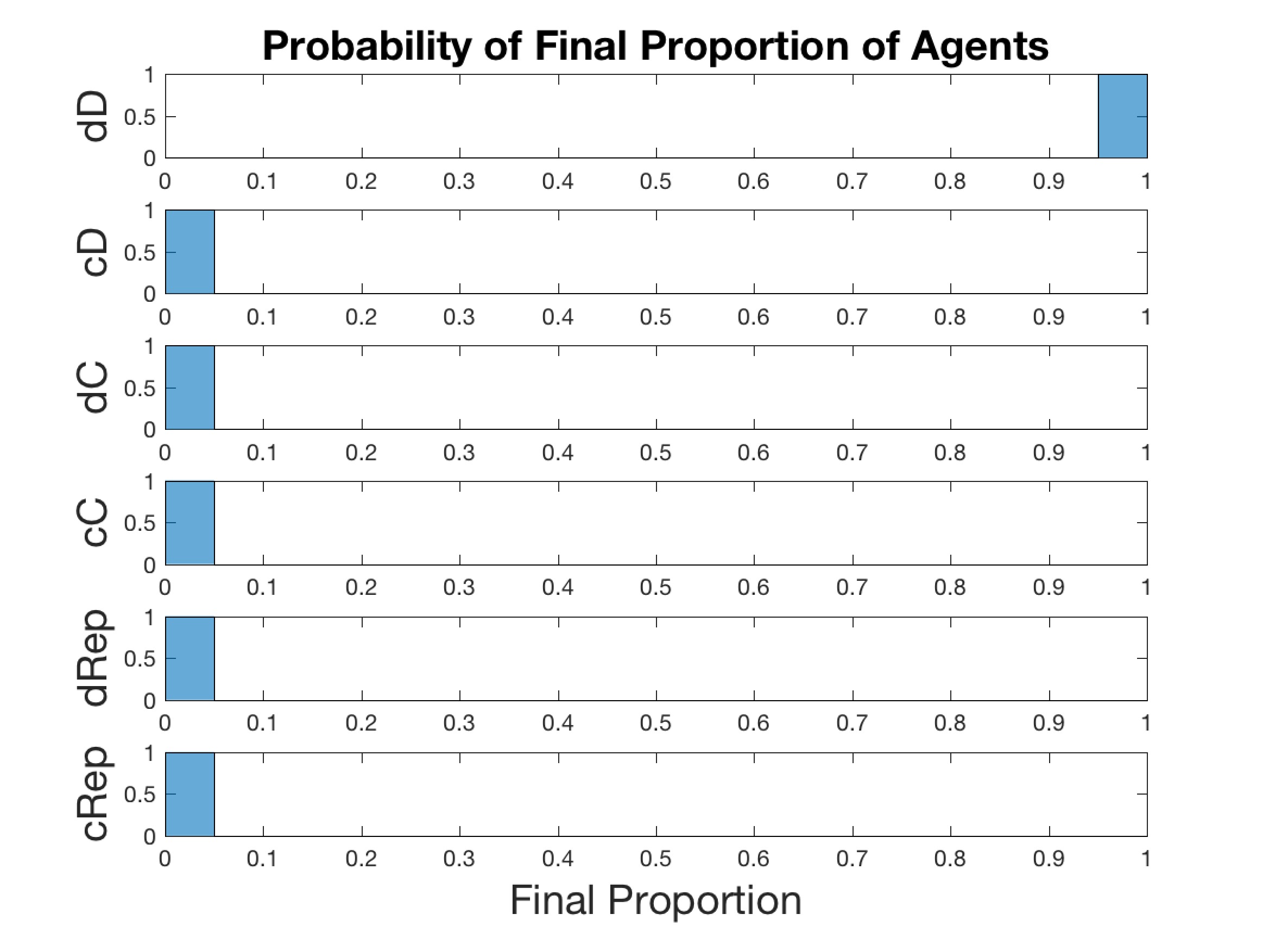}
\includegraphics[width=5cm]{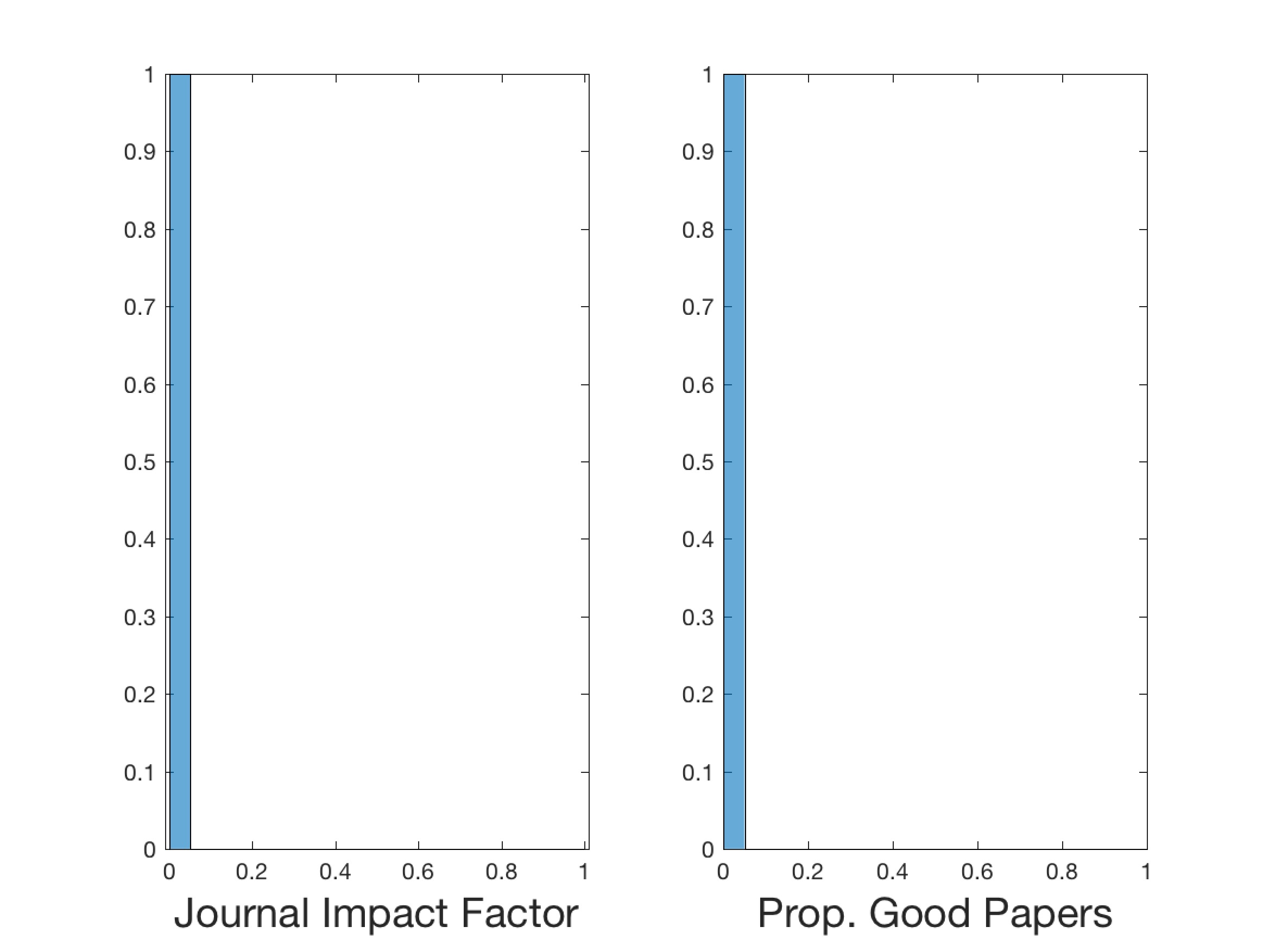}
\caption{Baseline distribution of each strategy type   (Left Panel)  and journal statistics (Right Panel) at the end of simulation. Results are distribution from 100 iterations with editorial strategy AP.  For all simulations: Review system is single blind; papers are assigned in random order to two random referees, the population is composed of 1200 individuals initially divided among the 6 possible bundled strategies; simulations run till convergence or until $t_{max}=100000$; papers are assigned in random order to randomly chosen referees; each referee can review up to 4 papers and a maximum of  30\% of the papers are publishable in each period.}
\label{Stats_baseline}
\end{figure}

We first demonstrate that in the baseline model in which scientists face costs for producing high-quality manuscripts and costs for producing high-quality reviews, there is no chance of any cooperation. This is not surprising, because low effort in writing papers as well as low effort in writing reviews is the dominant strategy in the baseline game. 
Figures \ref{Baseline_doubleblind}, \ref{Baseline_singleblind}, and \ref{Stats_baseline}
 report that this is the case both for double blind and for single blind peer review systems considering a neutral editorial policy (AP). In all cases, the strategy implying the production of low quality papers and reviews (dD) overtakes the entire population. To examine the failure of the scientific peer review process more closely, consider that high-quality review does not return any benefits in any case, therefore every reviewer is better off by choosing D. In a population with only dD and cD strategies, dD yields higher average payoffs since, if people review randomly, there is a 50\% chance of getting a low-quality paper published, which is exactly the same for high-quality submissions. As cD strategies do not benefit anything from peer review, but they entail higher costs for the author, they die out.
Without any feedback loop that would help to ensure the production of scientific quality, science ends up as an empty exercise.

\paragraph{Baseline with reputation-weighted consideration of reviews.}
Figure \ref{Baseline_doubleblind} and Figure \ref{Baseline_singleblind} contain also cases in which the editorial policy takes into account the editorial reputations of reviewers and attach higher weights to the opinion of higher reputed referees. These are editorial policies MR and ER. The strategy dD still gains overwhelming dominance under these policies.
This is because such policies do not ensure the occurrence of correlated equilibria: the larger weight to opinions does not mean at all that these opinions would favor high-quality contributions more than others. Low effort in reviewing is a dominant strategy for every author. The opinion of those who started with cooperation receive higher weights, but they still underscore defectors with regard to payoffs. The initial population that is
equally divided among different types of strategies goes through a quick and progressive elimination of cooperative strategies.
In this process, having a relatively bad reputation does not matter for payoffs as the chances of publishing a paper become equivalent for good and bad papers over time.

\paragraph{Baseline with the entry of reputational concerns.}
Somewhat surprisingly, the editorial policy that ensures a low, but stable level of cooperation is 1P. This is an editorial policy, which accepts all papers, if there is at least one positive recommendation from the reviewers. Seemingly, this policy is neutral to reputation. But in fact, this is not the case. As there is a constraint on how many papers can be published, the friendly 1P editorial policy generates the largest surplus of acceptable manuscripts. In case of a surplus, the editor ranks the papers based on the editorial reputation of the authors. Hence, a direct feedback on reputation exists, which is sufficient to guarantee in some simulation runs a stochastic mixed strategy equilibrium with the survival of cooperation.

\subsection{Reputation bias in editorial policy}

As we demonstrated, accounting on reputation of reviewers for making judgment on manuscripts is insufficient to trigger the production of high-quality reviews and high-quality papers. We should see if a more direct consideration of editorial reputation leads to higher efforts and as a result to better science. Note that the model extension in this direction is related intentionally to  popular discussions whether editors should be unbiased or they could rely on reputation signals of authors from the past. Should they catalyze the Matthew effect in science, in which the successful get even more success? Should they contribute to the maintenance of the old-boyism bias? If they do, does it hurt or help scientific development?  

Let us first introduce an editorial bias in favour of authors with high $REP_i^E$  and against authors with low one. We assume that authors with an editorial reputation higher than the median has a chance of desk acceptance that increases linearly with their reputation (Equation \ref{deskrejection}). Similarly, we assume that authors with an editorial reputation lower than the median has a chance of desk rejection that is in negative linear association with their reputation (Equation \ref{deskacceptance}).   
This modification does not rule out peer review, but concentrates its decisive character to the middle range, where no clear reputational judgment can be expected from the editor.

\begin{figure}
\centering
\includegraphics[width=5cm]{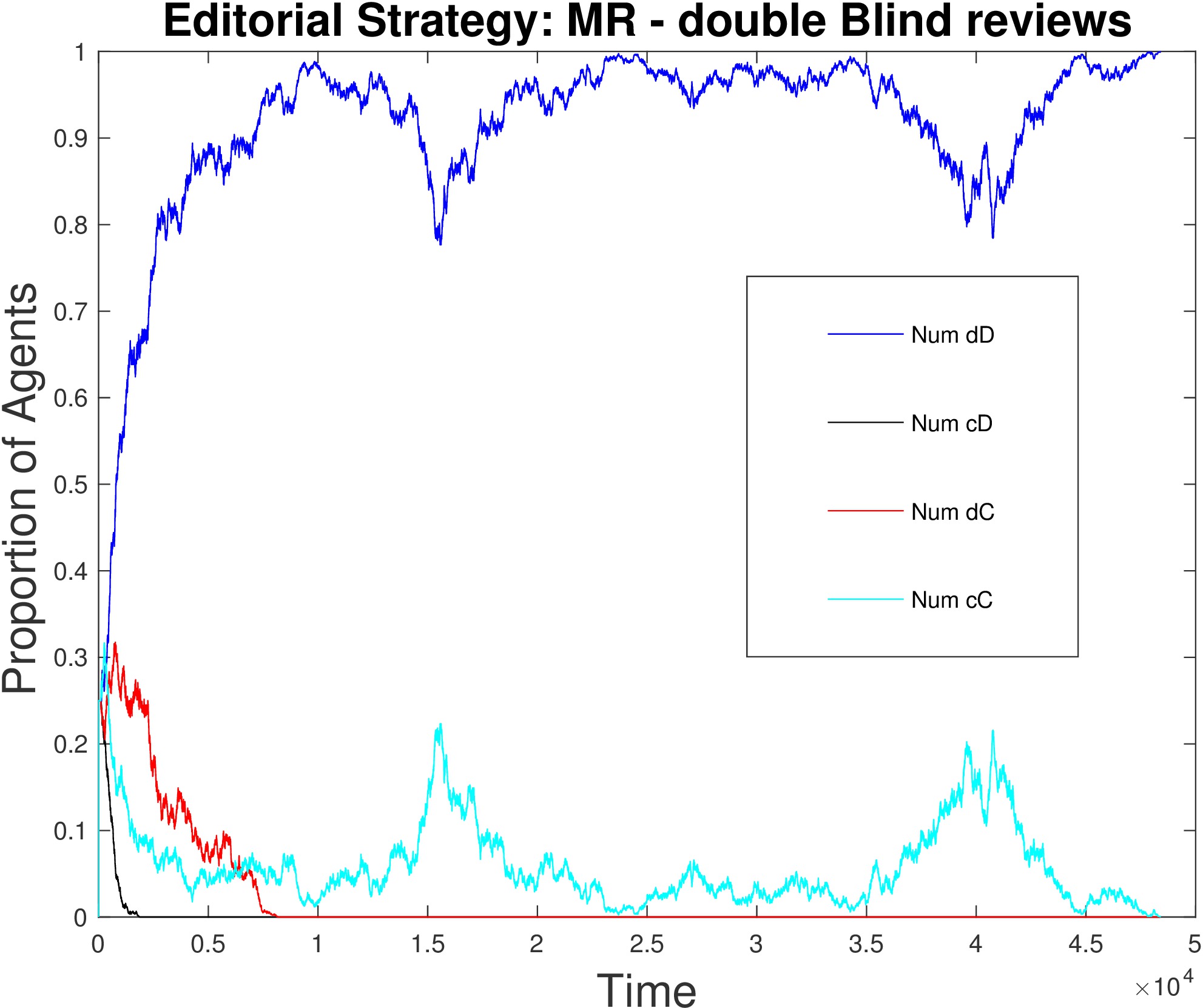}
\includegraphics[width=5cm]{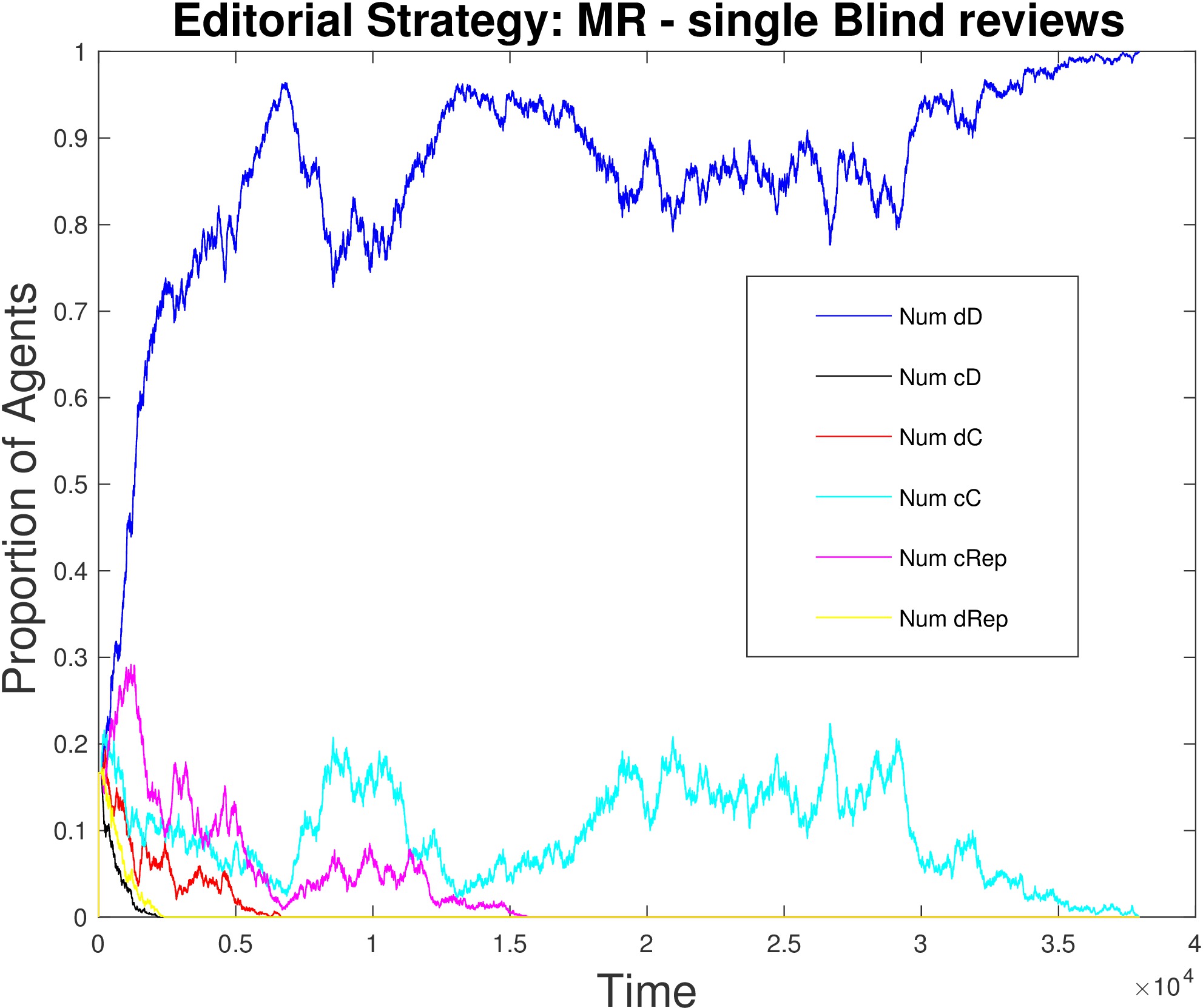}
\caption{The impact of reputation bias in editorial policy in desk acceptance and rejection of papers: double blind review (Left Panel) and single blind review (Right Panel). Papers are assigned in random order to two random referees. The editorial strategy is MR. Other parameters are as in the baseline.}
\label{ReputationEditorialAlone}
\end{figure}

\begin{figure}
\centering
\includegraphics[width=5cm]{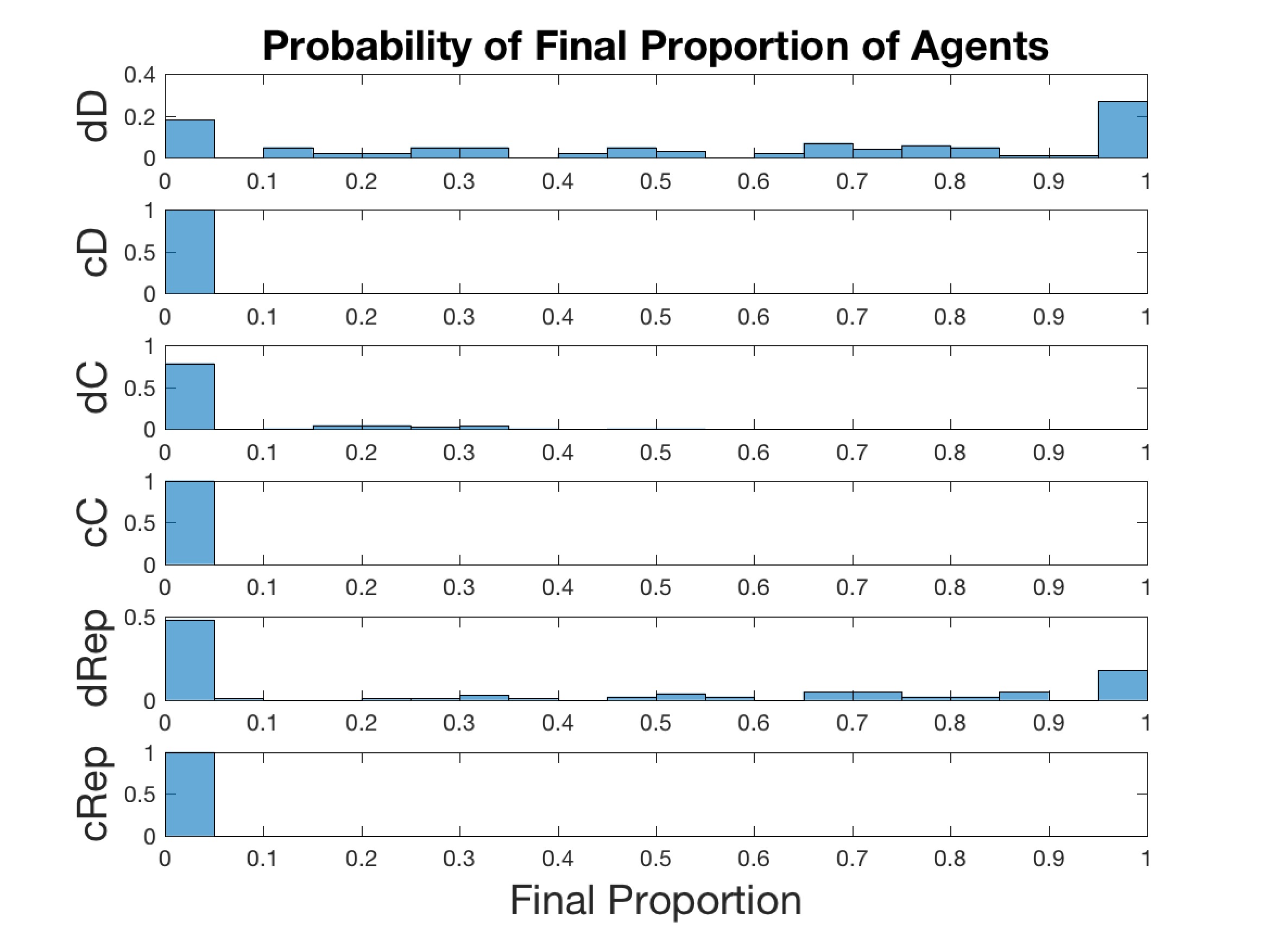}
\includegraphics[width=5cm]{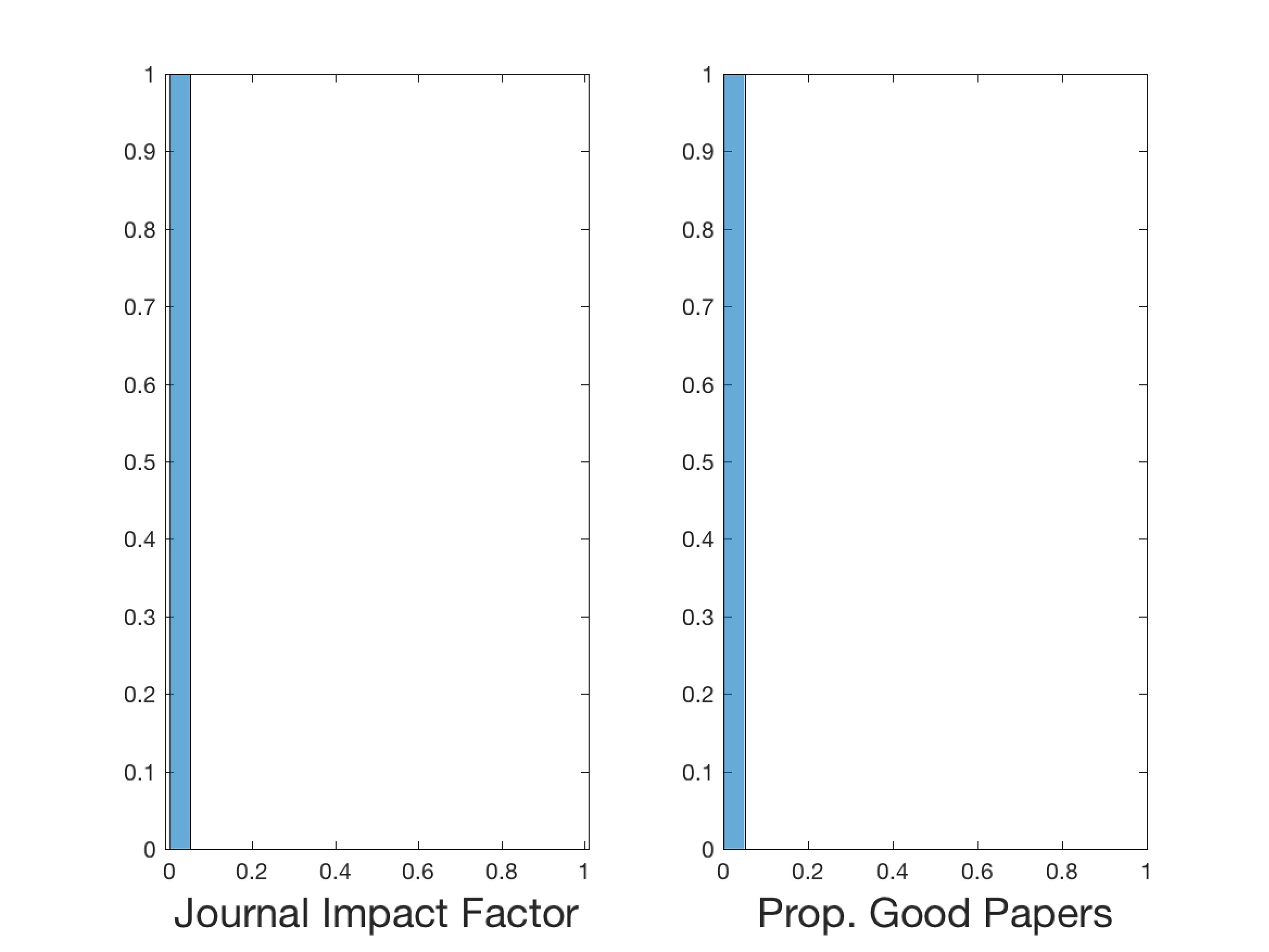}
\caption{The effect of using reputation in editorial policy. Distribution of each strategy type  (Left Panel)  and journal statistics (Right Panel) at the end of simulation. Results are distribution from 100 iterations with editorial strategy AP and single blind review system.  Other parameters are as in the baseline (See Fig \ref{Stats_baseline}).}
\label{Stats_REPUTATION}
\end{figure}

Results with this extension show no major breakthrough for cooperation: dD dominates the outcomes (Figure \ref{ReputationEditorialAlone}). 
Either with double blind or single blind peer review, all agents become of dD type.
This indicates that a direct editorial bias in desk acceptance and rejection in itself is insufficient to trigger a large extent of cooperation.
This kind of editorial bias, however, is able to support the survival of conditional cooperation of reviewers (Figure \ref{Stats_REPUTATION}).
When a strict editorial acceptance policy is applied (AP), the lack of publishable material leads to the need of selecting submissions based on reputation. The high effort in producing scientific material, however, does not pay off because of the difficulties of acceptance. Scientists therefore follow the easier path of gaining higher reputation and might place high effort in reviewing others. Reviewing efforts are profitable when they are most likely to provide reputation benefits. The public reputation helps the referees to get the best out of their reputation-based conditional strategy: when the public reputation of an author $REP^P_i$ is high, then it is more likely that his paper gets published, and therefore it is more likely that a review of high quality will ensure positive returns in terms of editorial reputation.
As a result, the cooperative reviewer strategy that is conditional on the reputation of the author might survive in case of single blind peer review. 
This happens because the editor might provide a differential treatment for individuals with higher reputations earned strictly by high-quality reviews.  
At the opposite, when the author's reputation is low, then reviewers with a strategy conditional on author reputation do not bother and follow the cheap strategy of providing random advice. In this case, their payoff is not different from agents who never put high effort in reviewing.

\subsection{Introducing the public good of Journal Impact Factor (JIF)}

\begin{figure}
\centering
\includegraphics[width=5cm]{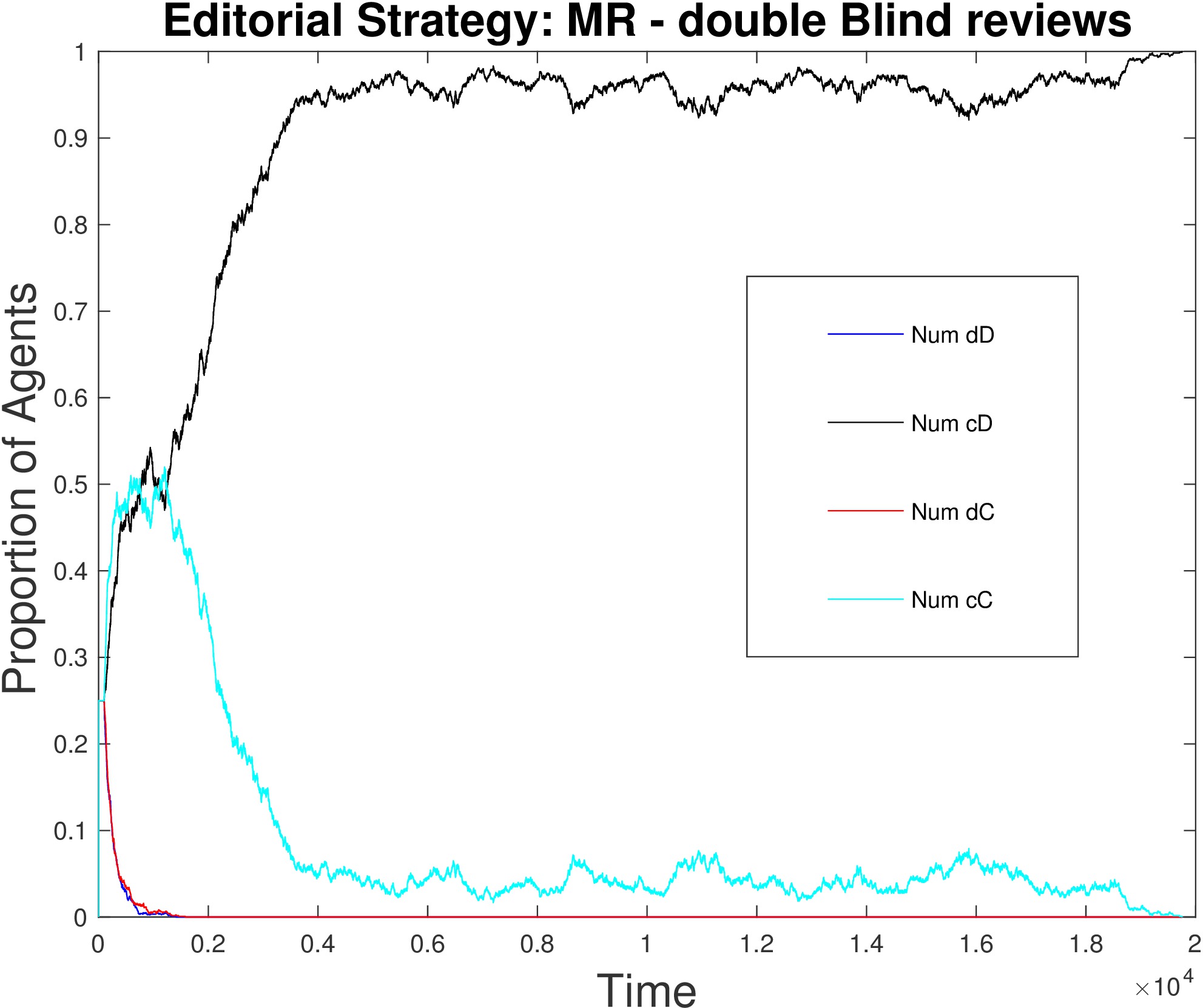}
\includegraphics[width=5cm]{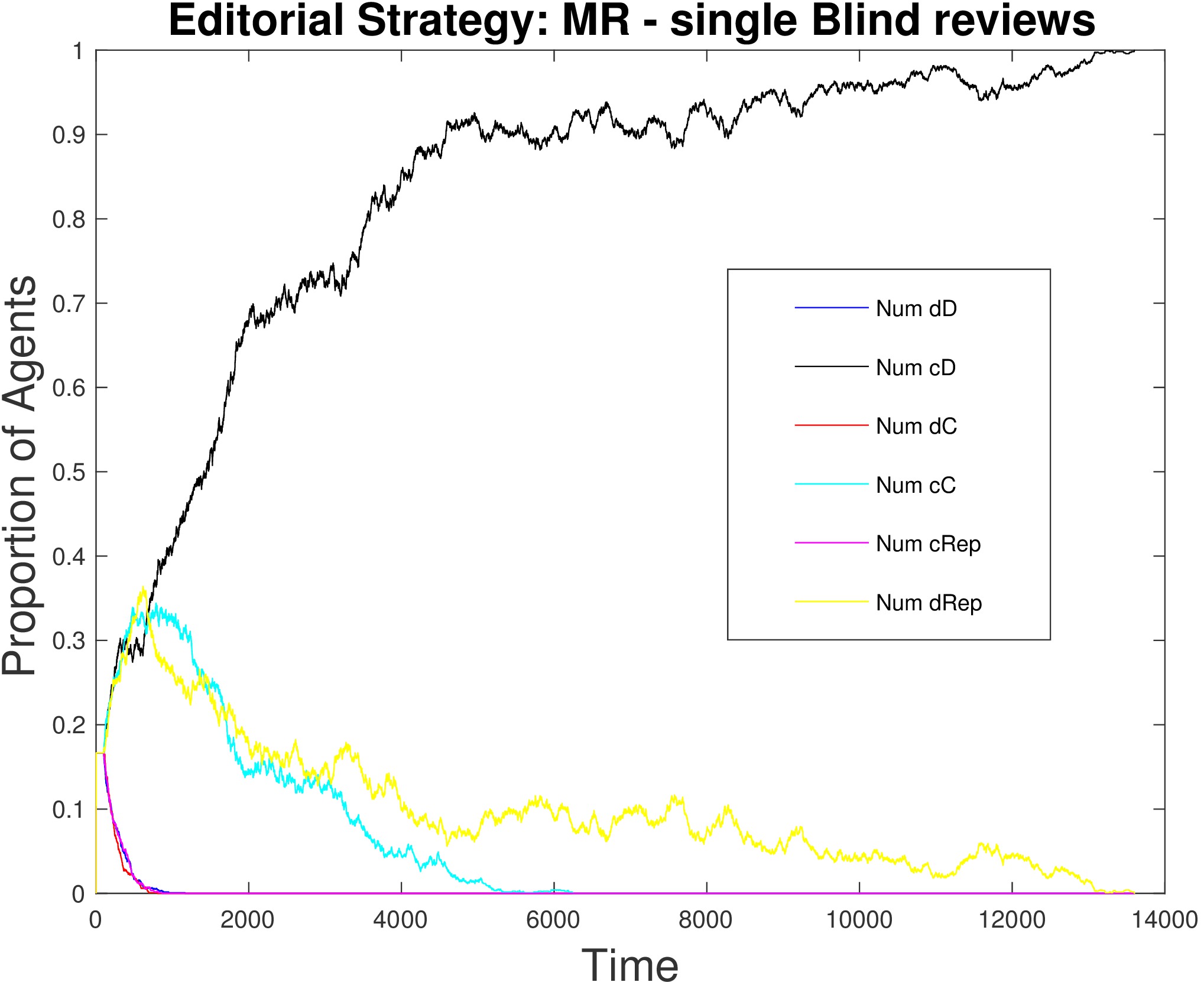}
\caption{The effect of introducing a Journal Impact Factor into the payoffs: in a double blind (Left Panel)  and single blind (right panel) peer review system. Papers are assigned in random order to two random referees.  Editorial strategy is MR. Other parameters are as in the baseline.}
\label{JIF_alone}
\end{figure}

The introduction of Journal Impact Factor implies that a public good bonus is added to the payoff of each agent publishing a paper. The size of the public good is proportional to the performance of the journal in terms of good papers published. Large public good benefits in the presence of some reputational motives allow for strategies producing high quality papers to survive and disseminate (Figure \ref{JIF_alone}). Furthermore, the analysis of the population evolution shows that when JIF is active, most strategies producing low-quality papers disappear from the population. This means that if a journal publishes high-quality papers, it ensures that submissions are also of high quality.

\begin{figure}
\centering
\includegraphics[width=5cm]{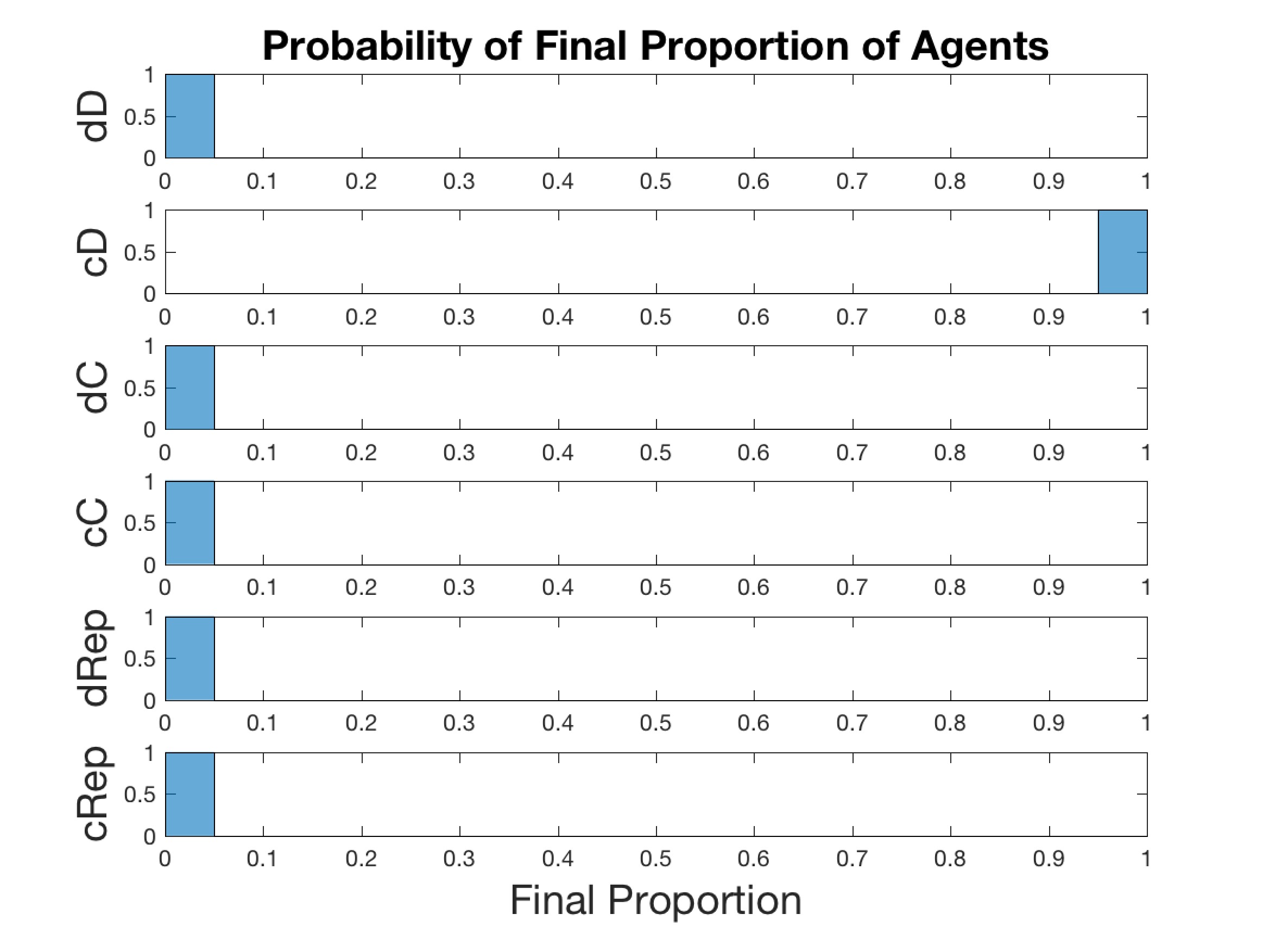}
\includegraphics[width=5cm]{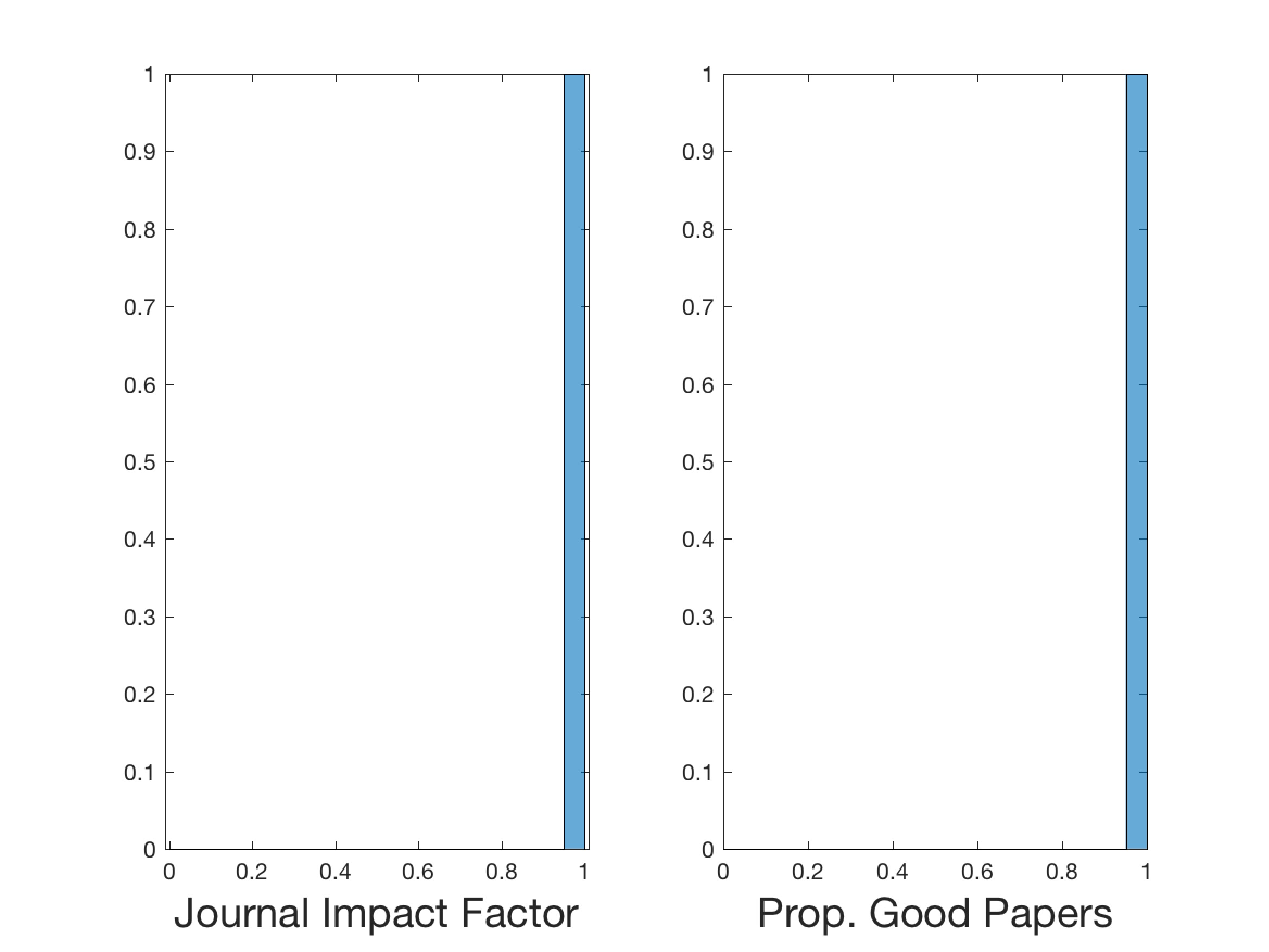}
\caption{The effect of accounting Journal Impact Factor into the payoffs. Distribution of each strategy type  (Left Panel)  and journal statistics (Right Panel) at the end of simulation. Results are distribution from 100 iterations with editorial strategy AP and single blind review system.  Other parameters are as in the baseline (See Fig \ref{Stats_baseline}).}
\label{Stats_JIF}
\end{figure}

This is good news given the fact that the public good reward of JIF as a payoff supplement does not erase the social dilemma structure of the game. 
Defection is still the best reply strategy both for authors and for reviewers. Still, cooperation evolves; thanks to the editorial account of author reputations and to the large initial share of cooperative strategies that survive the early phase of the simulation. Full cooperation is among those who disappear relatively late (Figure \ref{JIF_alone}), which assists the dominance of the  high-effort-in-writing and low-effort-in-reviewing strategy. As a consequence, the rise of good papers at the end is not accompanied by good reviews (Figure \ref{Stats_JIF}). 
Still, the scientific development is maintained at the best and results in the highest possible JIF. This means that only high quality papers are published. Peer review just adds a random noise for the publication process and it is meaningless anyway because everyone contributes with high-quality submissions. 

\paragraph{Journal Impact Factor together with strong reputation concerns.} 
Public good rewards that supplement the original payoff structure largely improve the opportunities for scientific development and lead to the overall success of the cD strategy. We have also seen that under certain editorial policies that take account of reputation, a low level of full cooperation (cC) can be sustained also without the public good reward. When we introduced desk rejections and acceptance based on reputation to the baseline, then some agents gained reputation successfully with a conditional reviewing strategy dRep. It is therefore interesting to observe which strategies are successful if both JIF and strong reputation concerns are accounted for. 

\begin{figure}
\centering
\includegraphics[width=5cm]{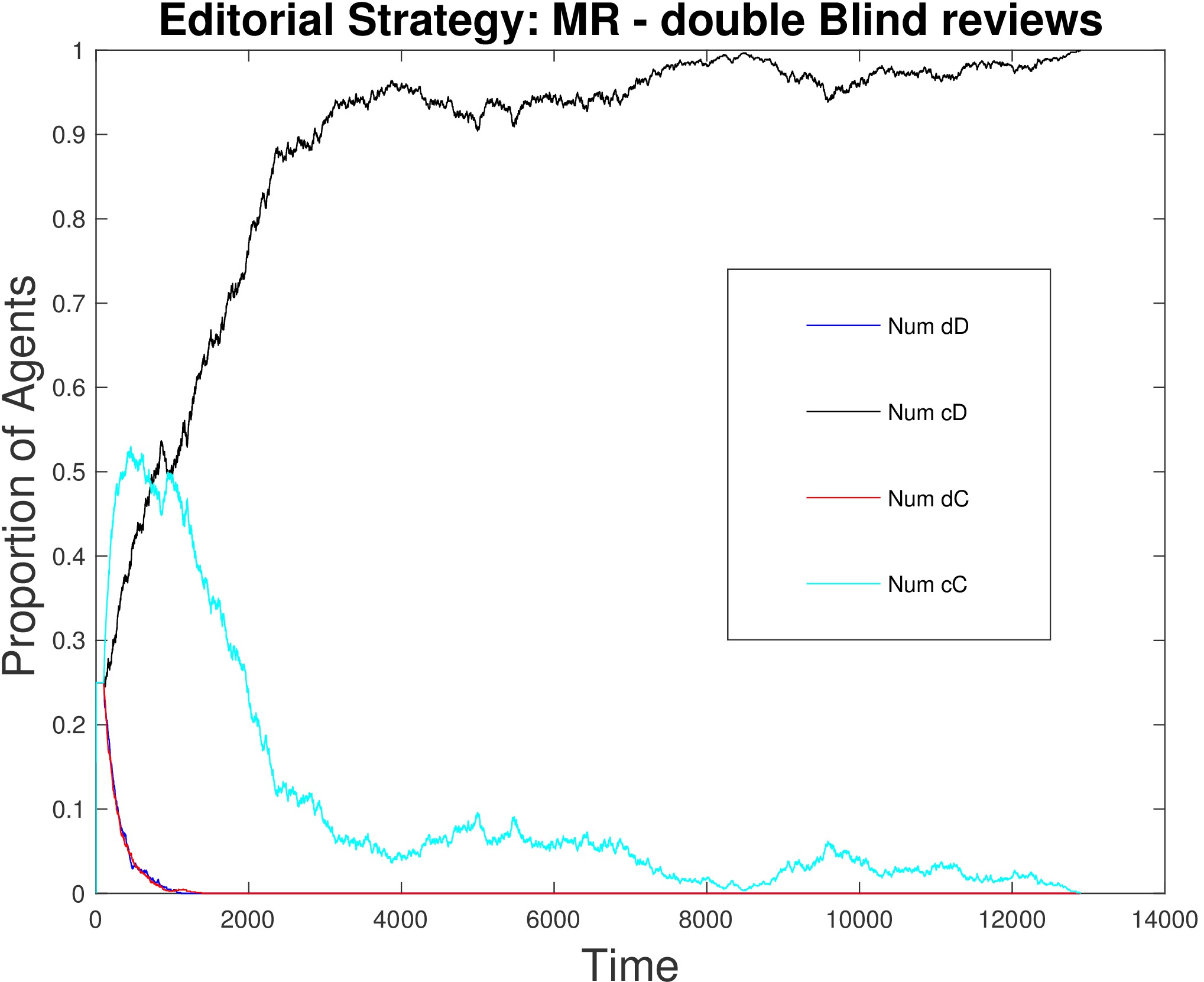}
\includegraphics[width=5cm]{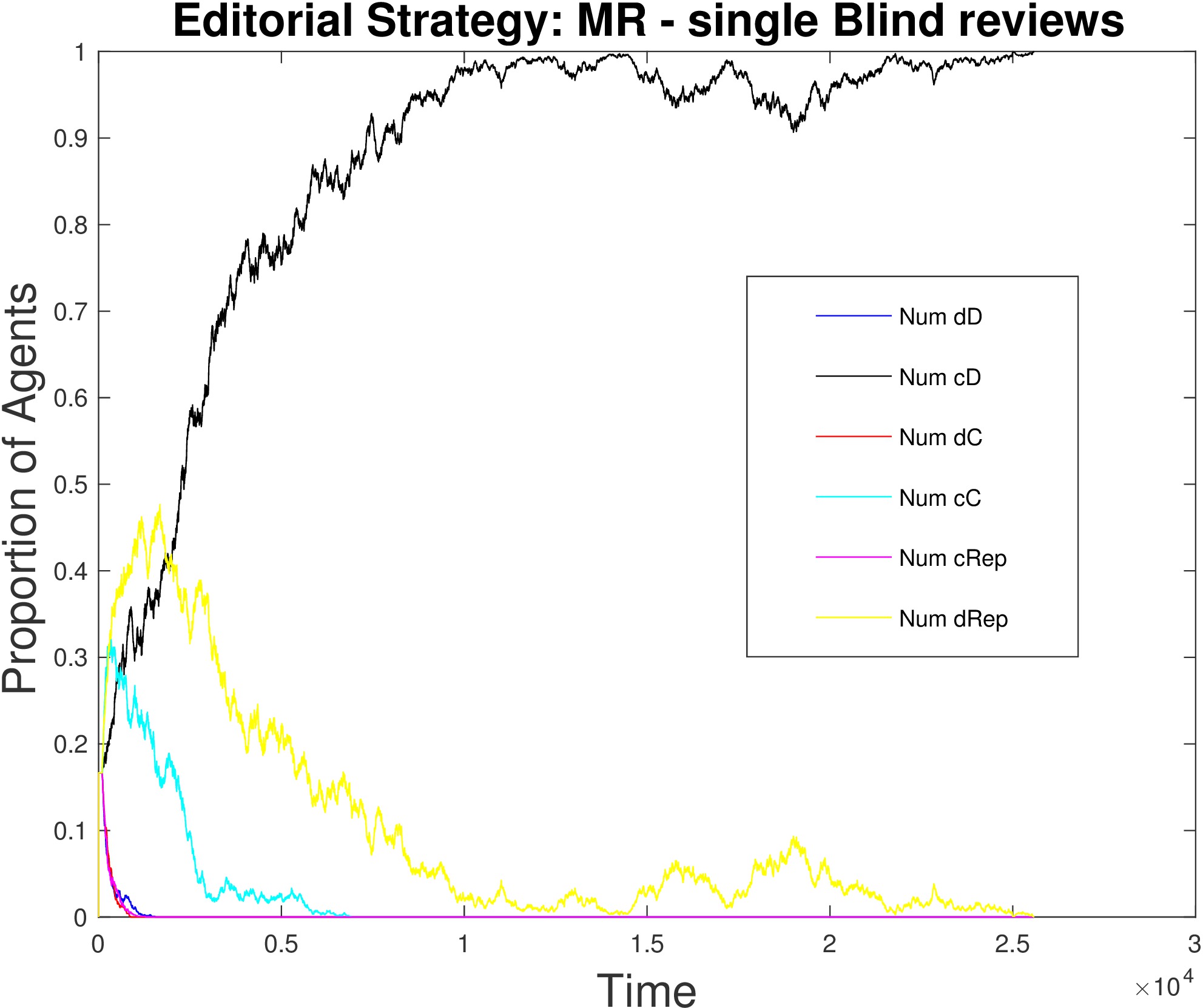}
\caption{The effect of accounting Journal Impact Factor into the payoffs and using  reputation in editorial policy: in a double blind (Left Panel)  and single blind (right panel) peer review system. Papers are assigned in random order to two random referees.  Editorial strategy is MR. Other parameters are as in the baseline.}
\label{JIFandReputation}
\end{figure}

\begin{figure}
\centering
\includegraphics[width=5cm]{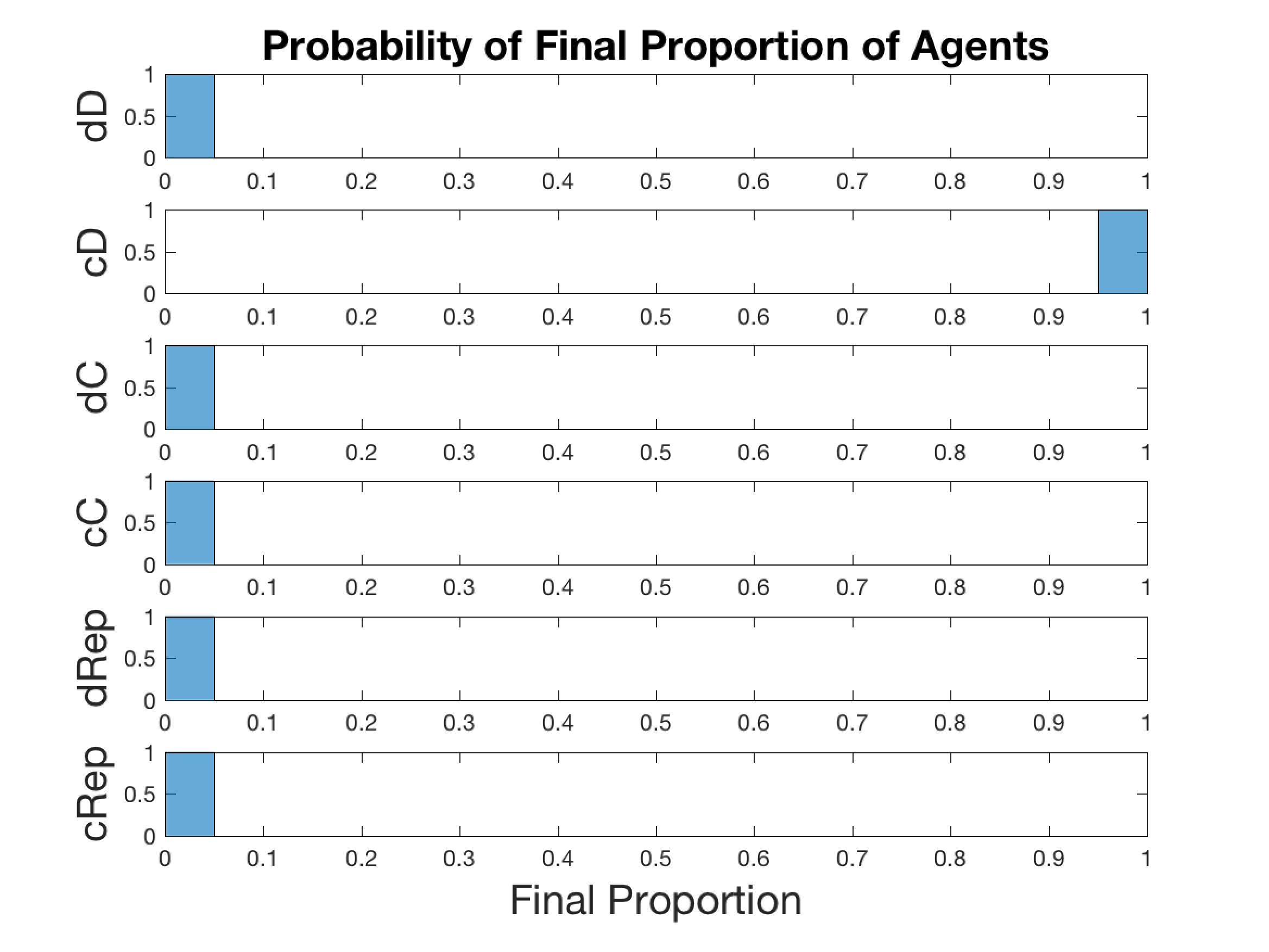}
\includegraphics[width=5cm]{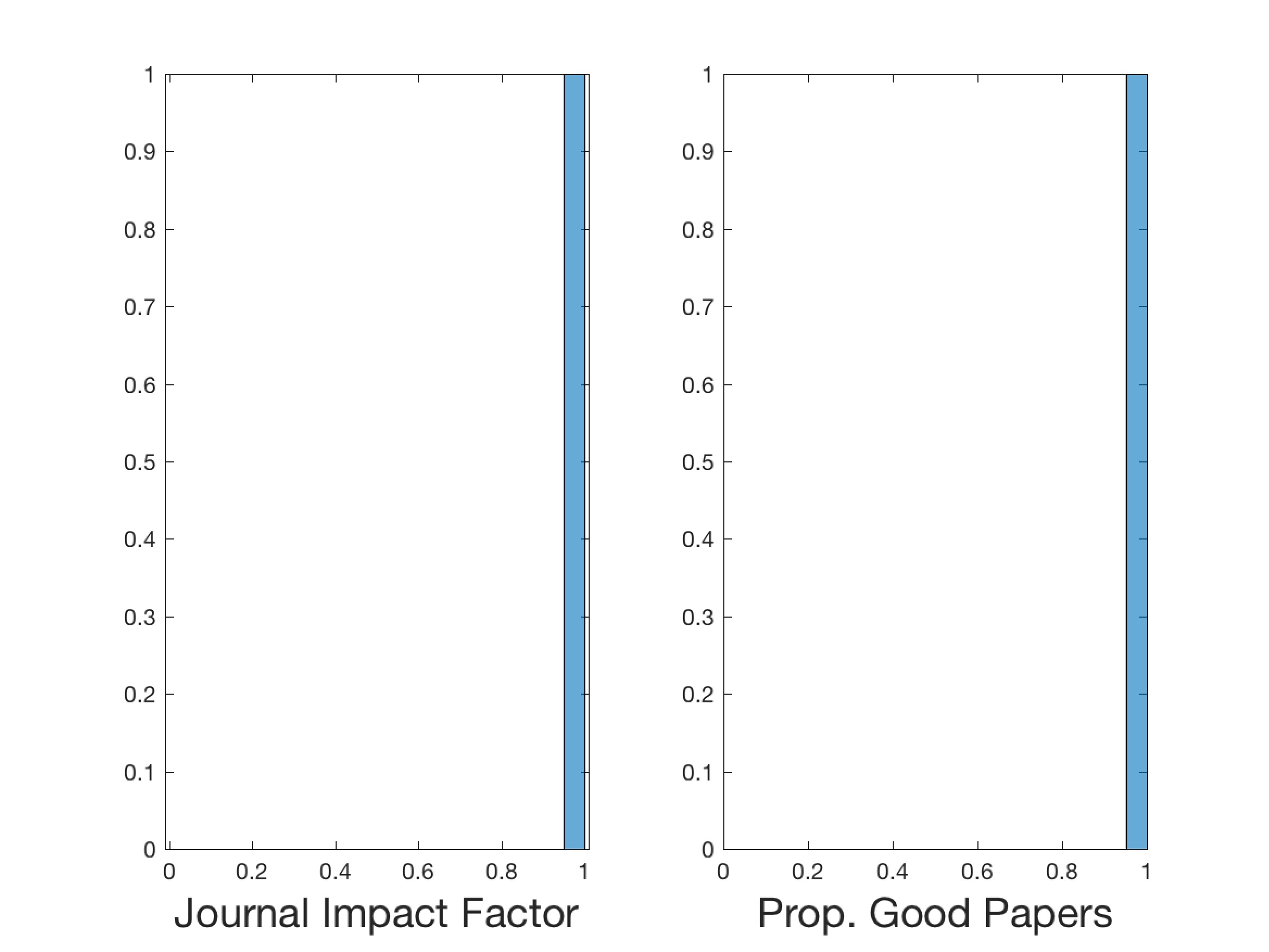}
\caption{The effect of accounting Journal Impact Factor into the payoffs and using reputation in editorial policy. Distribution of each strategy type (Left Panel) and journal statistics (Right Panel) at the end of simulation. Results are distribution from 100 iterations with editorial strategy AP and single blind review system. Other parameters are as in the baseline (See Fig \ref{Stats_baseline}).}
\label{Stats_JIFandREPUTATION}
\end{figure}

The results show that the strongest determinant of the evolution is the Journal Impact Factor (Figures \ref{JIFandReputation} and \ref{Stats_JIFandREPUTATION}). When it matters, even under strong reputation concerns, high-effort publication strategies gain dominance with low-effort reviews. This is very much meaningful once there is a reward for reputation and the most reputational gains can be obtained by high-quality publications.

\paragraph{Tacit agreements}
To complete the story, we still miss a mechanism that makes writing as well as reviewing papers plausible and sustainable without radically altering the payoff structure of the game. A realistic possibility is to consider "tacit agreements" that work against the impartiality of peer review in practice. Tacit agreements are based on direct reciprocity and might  work better in case of single blind reviews, in which at least one side of the informational asymmetry is relaxed. A "nice" tacit agreement strategy could start with high effort in the first $n$ rounds, recognizes previous reviewers of own papers with $p$ probability and retaliates them not with high effort/low effort, but with acceptance / rejection recommendation. Once there is a coordination device that brings high quality submissions in the hand of highly reputed reviewers, then acceptance recommendations would match true quality. In this case, the editorial selection policy will matter, because this would create the possibility of low cost reciprocation for high performing scientists.  This is in line with the conclusion of similar empirical and simulation work \cite{DondioetAlWP}. Many believe that this is the game that is played by the top researchers in top journals.
It is important to note, however, the possible drawbacks of self-emerged network-based practices, including old-boyism, partiality, and a conservative bias \cite{Sarigol2014,Sobkowicz2015,SooS2016}.


\section{Conclusion}
\label{conclusion}

It is puzzling why scientists devote considerable time and effort for writing reviews that decreases their time spent on their own research. Once everyone acts according to self-interest, reviews are all of low quality and they cannot be adequately used to judge scientific quality. As a consequence, scientists submit low quality work in the hope of passing through randomly judging reviewers. We have labeled this puzzle as the miracle of peer review and scientific development. We investigated some potential mechanisms that might resolve this puzzle by agent based modeling.
  
We have modeled scientific production accordingly: with an incentive structure in which low efforts in writing papers as well as in writing reviews is the dominant strategy of agents. We applied a replicator dynamics rule to the population of scientists, allowing for the reproduction of strategies that result in higher payoffs. Not surprisingly, low effort in writing papers as well as in writing reviews have spread in the population and scientific practice has become an empty exercise in our baseline model.

Next, we assumed that editors might rely on the reputations of authors in their choices. In our model with a single journal, the editor took perfect account of high- and low-quality publications of authors, the number of their rejected papers, and if their reviewer recommendations were in line with the true quality of the paper or not. We examined different editorial policies that took account of the reputations of scientists. 
We showed that if reputations are used in reviewer selection, then it does not save science from low quality submissions and low quality reviews. A bit more surprisingly, easing the route for publication by desk acceptance for highly reputed authors alone has not changed anything either. All this indicated that the emergence of cooperation in the form of high efforts is an extremely difficult puzzle.

Some cooperation has resulted from a friendly editorial policy that categorized submissions as publishable if at least one reviewer recommended publication. This policy led to an oversupply of publishable material, which called for the ranking of submissions based on author reputations. This direct feedback has made the investment in reputations profitable. Consequently, high-effort author strategies survived in a mixed equilibrium together with low-effort author strategies, but nobody invested effort in reviewing.

Also when a strict editorial acceptance policy was applied, in which only papers with unanimous reviewer support are published, some cooperation has emerged. In this case, however, the lack of publishable material was responsible for the worth of reputation. As the investment in reputation via writing papers was risky due to the difficulties of publishing, scientists profited more from the investment in reputation via reviews. A strategy that conditioned high reviewing efforts on the author's reputation was able to gain a notable share in the population.

Reputations worked to some extent, but public good benefits worked clearly better for scientific development. Once we introduced the Journal Impact Factor as a public good benefit, which meant the distribution of an additional payoff for all authors who published in the journal (either a good or a bad paper), cooperation has become the most successful strategy of authors. In this case, editorial reputations became correlated with actual contributions to the provision of the pubic good. But as the production of high quality papers was still much more important for reputations than high quality reviews, the cooperative strategy that emerged as successful was investing high effort only in manuscript writing and not in reviewing. As a consequence, cooperation has been observed in scientific production, but peer review has just added random noise to this development, which raises doubts of its use \cite{Neff2006} and concerns about the use of public money.

At the end, we were successful in demonstrating in a simple model the puzzling motivational problem of peer review. We highlighted that it is not easy to find the way out of this puzzle. We showed that a high-value of the public good of science maintains scientific development. Reputational systems that are heavily building on author contributions might be partially sucessful, especially if the reputational hierarchy is directly used for selecting between similarly rated submissions. These mechanisms, however, will not help to sustain the efficiency of peer review. Paradoxically, mechanisms that are able to induce some level of high-quality reviews are building on reciprocity, and in practice they are often associated with impartiality, old-boyism, the emergence of invisible colleges, the Matthew effect, the conservative bias, and the stratification of science.

We plan to extend our simple model towards studying multiple journals that compete for success with each other. This extension allows for the evolution of editorial strategies in a straightforward way and in parallel to theoretical studies that highlight how group selection can ensure higher cooperation \cite{Traulsen2006,Nowak2006}, is expected to lead to better reviewer performance.


\begin{thebibliography}{99}

\bibitem{Alberts2008} Alberts, Bruce, Brooks Hanson, and Katrina L. Kelner. "Reviewing peer review." Science 321.5885 (2008): 15-15.

\bibitem{Axelrod1984} Axelrod, Robert. "The evolution of cooperation.", New York, Basic Books (1984).

\bibitem{Axelrod1981} Axelrod, Robert, and William Donald Hamilton. "The evolution of cooperation." Science 211.4489 (1981): 1390-1396.

\bibitem{Barrera2008} Barrera, Davide. "The social mechanisms of trust." Sociologica 2.2 (2008). Doi: 10.2383/27728.

\bibitem{Barrera2009} Barrera, D. and Buskens, V. 2009 "Third-Party Effects on Trust in an Embedded Investment Game." In Cook, K., Snijders, C., Buskens, V. And Cheshire (eds), Trust and Reputation, New York, Russell Sage, 37-72. 

\bibitem{Bernstein2013} Bernstein, Joseph. "Free for service: the inadequate incentives for quality peer review." Clinical orthopaedics and related research 471.10 (2013): 3093.

\bibitem{BianchiEtAlWp} Bianchi, Federico, Francisco Grimaldo, Lorena Cadavid, Giangiacomo Bravo, and Flaminio Squazzoni. 2016. "Peer review as a cooperation dilemma: An agent-based model." Work in progress.

\bibitem{Bornmann2013} Bornmann, Lutz. "Evaluations by peer review in science." Springer Science Reviews 1.1-2 (2013): 1-4.

\bibitem{Boyd1989} Boyd, Robert, and Peter J. Richerson. "The evolution of indirect reciprocity." Social Networks 11.3 (1989): 213-236.

\bibitem{BravoSquazzoniTakacs} Bravo, Giangiacomo, Flaminio Squazzoni, and K\'{a}roly Tak\'{a}cs. "Intermediaries in Trust: Indirect Reciprocity, Incentives, and Norms." Journal of Applied Mathematics 2015 (2015).

\bibitem{Chetty2014} Chetty, R., Saez, E., and S\'{a}ndor, L. "What Policies Increase Prosocial Behavior? An Experiment with Referees at the Journal of Public Economics." Journal of Economic Perspectives, (2014) 28(3): 169-188.

\bibitem{Coleman1986} Coleman, James S. "Social structure and the emergence of norms among rational actors." Paradoxical Effects of Social Behavior. Physica-Verlag HD, 1986. 55-83.

\bibitem{Dellarocas2003} Dellarocas, C. 2003. Management Science, 49: 1407-1424. 

\bibitem{DondioetAlWP} Dondio, Pierpaolo, Niccol\'{o} Casnici, Francisco Grimaldo, Flaminio Squazzoni, and John Kelleher 2016. Work in progress.

\bibitem{Gilbert1997} Gilbert, Nigel. 1997. "A Simulation of the Structure of Academic Science." Sociological Research Online,  2 (2), http://www.socresonline.org.uk/socresonline/2/2/3.html.

\bibitem{Malicki} Malicki, Mario and Bahar Mehmani 2016. Motivations for reviewing manuscripts submitted to Elsevier journals. Work in progress.

\bibitem{Milinski2002_1} Milinski, Manfred, Dirk Semmann, and Hans-J\:{u}rgen Krambeck. "Reputation helps solve the 'tragedy of the commons'." Nature 415.6870 (2002): 424-426.

\bibitem{Milinski_2002a} Milinski, Manfred, Dirk Semmann, and H. Krambeck. "Donors to charity gain in both indirect reciprocity and political reputation." Proceedings of the Royal Society of London B: Biological Sciences 269.1494 (2002): 881-883.

\bibitem{Milinski_2001}Milinski, Manfred, et al. "Cooperation through indirect reciprocity: image scoring or standing strategy?" Proceedings of the Royal Society of London B: Biological Sciences 268.1484 (2001): 2495-2501.

\bibitem{Neff2006} Neff, B. D. and Olden, J. D. 2006. Is Peer Review a Game of Chance? BioScience, 56(4):333-340.

\bibitem{Nowak2006} Nowak, Martin A. "Five rules for the evolution of cooperation." Science 314.5805 (2006): 1560-1563.

\bibitem{Nowak2005} Nowak, Martin A. and Karl Sigmund. "Evolution of indirect reciprocity." Nature 437.7063 (2005): 1291-1298.

\bibitem{Olson1965} Olson, M. Jr. "The Logic of Collective Action". Cambridge (Mass.), Harvard University Press  (1965).

\bibitem{Paolucci2014} Paolucci, Mario and Jaime Sim\~{a}o Sichman. "Reputation to understand society." Computational and Mathematical Organization Theory 20.2 (2014): 211.

\bibitem{PaolucciGrimaldo2014} Paolucci, Mario and Francisco Grimaldo. 2014. "Mechanism change in the simulation of peer review." Scientometrics doi 10.1007/s11192-014-1239-1.

\bibitem{Sarigol2014} Sarigol, Emre, et al. "Predicting scientific success based on coauthorship networks." EPJ Data Science 3.1 (2014): 1-16.

\bibitem{Seeber2016} Seeber, Marco and Alberto Bacchelli 2016. "Does single blind peer review hinder newcomers?", Work in progress.

\bibitem{Semmann2005} Semmann, Dirk, Hans-J\:{u}rgen Krambeck, and Manfred Milinski. "Reputation is valuable within and outside one's own social group." Behavioral Ecology and Sociobiology 57.6 (2005): 611-616.

\bibitem{Sobkowicz2015} Sobkowicz, Pawel. "Innovation Suppression and Clique Evolution in Peer-Review-Based, Competitive Research Funding Systems: An Agent-Based Model." Journal of Artificial Societies and Social Simulation 18.2 (2015): 13.

\bibitem{Sommerfeld2007} Sommerfeld, Ralf D., et al. "Gossip as an alternative for direct observation in games of indirect reciprocity." Proceedings of the National Academy of Sciences 104.44 (2007): 17435-17440.

\bibitem{Sommerfeld} Sommerfeld, Ralf D., Hans-Juergen Krambeck, and Manfred Milinski. "Multiple gossip statements and their effect on reputation and trustworthiness." Proceedings of the Royal Society of London B: Biological Sciences 275.1650 (2008): 2529-2536.

\bibitem{SooS2016} So\'{o}s, S\'{a}ndor,  Zs\'{o}fia Vida, Beatriz Barros, Ricardo Conejo, Richard Walker. Social networks as a potential source of bias in peer review (2016) Working paper.

\bibitem{SquazzoniTakacs2011} Squazzoni, Flaminio, and K\'{a}roly Tak\'{a}cs. "Social simulation that'peers into peer review'." Journal of Artificial Societies and Social Simulation 14.4 (2011): 3.

\bibitem{SquazzoniBravoTakacs2013} Squazzoni, Flaminio, Giangiacomo Bravo, and K\'{a}roly Tak\'{a}cs. "Does incentive provision increase the quality of peer review? An experimental study." Research Policy 42.1 (2013): 287-294.

\bibitem{Traulsen2006} Traulsen, Arne and Martin A. Nowak. "Evolution of cooperation by multilevel selection." Proceedings of the National Academy of Sciences 103.29 (2006): 10952-10955.

\bibitem{Warne2016} Warne, Verity. "Rewarding reviewers-sense or sensibility? A Wiley study explained." Learned Publishing 29.1 (2016): 41-50.


\end{thebibliography}
\end{document}